\documentclass[useAMS,usegraphicx,usenatbib]{mn2e}
\usepackage{amsmath}
\usepackage{amsfonts}
\usepackage{amssymb}
\usepackage{verbatim}
\usepackage{mathrsfs}
\usepackage{enumerate} 
\usepackage{paralist}
\newif\ifAMStwofonts
\AMStwofontstrue
\title [GRB host galaxies] {Chemical evolution models: GRB host identification and cosmic dust predictions}
\author[V. Grieco et al.] {V. Grieco,$^{1,3}$\thanks{E-mail:grieco@oats.inaf.it} F. Matteucci,$^{1,2,3}$  F. Calura,$^{4}$  
S. Boissier,$^{5,4}$ F.  Longo, $^{1,3}$ and V. D'Elia, $^{6,7}$\\
$^{1}$Dipartimento di Fisica, Sezione di Astronomia, Universit\`a di Trieste, via G.B. Tiepolo 11, I-34131, Trieste, Italy \\
$^{2}$I.N.A.F., Osservatorio Astronomico di Trieste, via G.B. Tiepolo 11, I-34131, Trieste, Italy\\
$^{3}$I.N.F.N., Trieste, Via A. Valerio 2, I-34127 Trieste, Italy\\
$^{4}$I.N.A.F., Osservatorio Astronomico di Bologna, via Ranzani 1 I-40127 Bologna, Italy \\
$^{5}$ Aix Marseille Universit\'e, CNRS, LAM (Laboratoire d'Astrophysique de Marseille) UMR 7326, 13388, Marseille\\
$^{6}$I.N.A.F., Osservatorio Astronomico di Roma, via Frascati 33, 00040, Monteporzio Catone, Italy \\
$^{7}$A.S.I., Science Data Center, via del Politecnico snc, 00133, Rome, Italy}

\begin{document}
\date{Accepted . ; in original form 2014}

\pagerange{\pageref{firstpage}--\pageref{lastpage}} \pubyear{2002}

\maketitle

\label{firstpage}

\begin{abstract}

The nature of some GRB host galaxies has been investigated by means of chemical evolution models of galaxies of different morphological type 
following  the evolution of the abundances of H, He, C, N, O, $\alpha$-elements, Ni, Fe, Zn, and including also the evolution of dust. 
By comparing predictions with abundance data, we were able to constrain nature and age of GRB hosts. 
We also computed a theoretical cosmic dust rate, including stellar dust production, accretion and destruction,
under the hypotheses of pure luminosity evolution and strong number density evolution of galaxies. 
We suggest that one of the three GRB hosts is a massive proto-spheroid catched during its  formation, 
while for the other two the situation
is more uncertain, although one could perhaps be a spheroid and the other a spiral galaxy. 
We estimated the chemical ages of the host galaxies which vary from 15 to 320 Myr.
Concerning the cosmic effective dust production rate in an unitary volume of the Universe, our results show that in the case of pure 
luminosity evolution there is a first peak between redshift z=8 and 9  and another at $z\sim 5$, whereas in the case of strong number 
density evolution it increases slightly from z=10 to $z\sim 2$ and then it decreases down to z=0. Finally, we found tha the total cosmic 
dust mass density at the present time is:  $\Omega_{dust} \sim 3.5\cdot 10^{-5}$in the case of pure luminosity evolution and  
$\Omega_{dust} \sim 7\cdot 10^{-5}$ in the case of number density evolution.

\end{abstract}


\section{Introduction}

Most LGRBs are associated with massive stellar explosions and consequently they are linked to ongoing star formation.
Savaglio et al. (2009), studying a sample of GRB host galaxies (HGs), concluded that  there is no clear evidence that 
HGs are peculiar galaxies and that they are instead similar to normal star-forming galaxies, both in the local and in the
early universe. For this reason, understanding the nature of HGs provides constraints on the GRBs progenitors:  
the collapsar model (Woosley 1993; Woosley \& Heger 2006) predicts a low metallicity environment, typical of dwarf
galaxies, because of the reduced angular momentum loss and mass loss via strong winds at the surface of the star,
while in other progenitor models there is no such an assumption (e.g. Ouyed et al. 2005; Fryer \& Heger 2005).

Many attempts have been made to characterize the hosts of GRBs (e.g. Fruchter et al. (1999,2006), Castro Cer\'on et al. (2006, 2010), 
Le Floc’h et al. (2006), Modjaz et al. (2008), Levesque et al. (2010), Perley et al. (2013) and 
Boissier et al. (2013).
In particular, Berger et al. (2003) and  Le Floc’h et al. (2006), 
using mid-IR and radio observations, showed that some GRB host galaxies are 
characterized by a large amount of star formation obscured by dust and in general they have attempted to understand the properties of GRB hosts, 
but only few of them have used powerful chemical evolution indicators as we are going to do in this work. 
Moreover, the importance of dust is also confirmed by the
large fraction of GRBs occurring in dusty environments (e.g. Perley et al. 2009, 2013; Greiner et al. 2011; Hatsukade et al. 2012).

Dust is present in almost every astrophysical environment affecting the comprehension of their physical properties, 
such as their star formation rate and abundance ratios. Dust grains are made up of metals, primarily C, O, Mg, Si, 
Fe, Ni, S, Ca which enriched the interstellar medium (ISM) during particular stellar evolutionary phases  (e.g. asymptotic giant branch (AGB) stars and 
core collapse supernovae. 

Interesting dust evolution studies in galaxies have been performed by Dwek (1998), Dwek et al. (2007, 2011), Calura et al. (2008) and Schurer et al. (2009).
In those papers, models of galactic chemical evolution including dust were presented. 
The basic idea of identifying GRB host galaxies 
by means of chemical evolution models is based on the fact that galaxies of different morphological type should show different abundance patterns. 
Such different patterns derive from their different star formation histories (SFHs). Calura et al. (2009) presented a first approach were the SFHs of the HGs of GRBs
were derived by comparison between observed abundances and model results; to do that they adopted the Calura et al. (2008) models including dust evolution. 
In the following years, Fan et al. (2010) also deduced the nature of some hosts of GRBs  using chemical models but the dust was not considered.
Both those papers concluded that the hosts of the studied GRBs were probably irregular galaxies.
Clearly, the presence of dust can affect the abundances derived from the gas in galaxies and therefore it is very important to include the possibility of 
dust condensation in chemical evolution models.

In this paper, by adopting the dust evolution as in  Calura et al. (2008), we aim at constraining the nature and SFH of some HGs by means of detailed 
chemical evolution models taking into account the variation in the abundance patterns. 
Moreover, we aim at computing the dust evolution in each type of galaxy, including processes of stellar dust production, accretion of dust onto preexisting grains 
and dust destruction. As a byproduct,  we are going to compute the cosmic effective dust production rate in an unitary volume of the Universe by taking into account 
the contributions from all galaxies. 

Section \ref{models} describes the chemical evolution models used in this work and the nucleosynthesis prescriptions; in section \ref{dust}
the prescriptions for the supernova rates and the dust production, accretion and destruction rates in galaxies are presented. In section \ref{sample} 
we briefly show the observational data. 
In section \ref{res}  the results for dust production, accretion and destruction in galaxies of different morphological type  as well the results deriving from the 
comparison between our theoretical predictions and observational abundance data of GRB 081008,120327A and 120815 host galaxies are shown. 
The definition and computation of the cosmic effective dust production rates, in different frameworks of galaxy formation, 
are presented in Section \ref{cosmic}, together with the calculation of the present time cosmic dust density.
Finally, in section \ref{concl}, some conclusions are drawn.

\section[]{The chemical evolution models} \label{models}

We follow the evolution of the abundances of several chemical species and of 
the dust content of spirals, elliptical and irregular galaxies using  detailed 
self-consistent chemical evolution models able to reproduce the majority of the 
properties of the different types of galaxies. In all models the instantaneous 
recycling approximation is relaxed and the stellar lifetimes are taken into account. The models all assume that galaxies form by gas accretion which accumulates a final mass called infall mass ($M_{tot}$).

The model takes into account the feedback from SNe and stellar winds 
giving rise to a galactic outﬂow when the gas thermal energy exceeds the binding energy of gas.

In our models the SFR is a simple Schmidt’s (1963) law:
\begin{equation}
 \psi(t)=\nu G(t)
\end{equation}
where the quantity $\nu$ is the star formation efficiency (SFE), namely the inverse of the typical time-scale for star formation, expressed in $\mbox{Gyr}^{-1}$,
and it has been varied as a function of the galaxy morphological type (see Section 5). $G(t)$ is the gas mass fraction expressed as $G(t)=M_{gas}(t)/M_{tot} $, the ratio between the gas mass and the infall mass. Dividing by the infall mass it is only for a normalization purpose and all the other variables are also normalized in the same way. Because of this, $\psi(t)= \dot M_{gas}(t)/ M_{tot}$.
All the models used in this paper consider only one gas phase.
The chemical evolution equation for a given chemical element i takes the following form:
\begin{equation} \label{chem-ISM}
\dot {G}_{i}(t) = -\psi(t)X_i(t) + R_i(t) + \dot {G}_{i,inf}(t) - \dot {G}_{i,w}(t) 
\end{equation}
where $G_{i}(t)=M_{gas}(t)X_{i}(t)/M_{tot}=G(t)X_{i}(t)$
is the fractional mass of the element \emph{i} at the time \emph{t} in the ISM.
, and $X_{i}(t)$ represents the mass fraction of element \emph{i}
in the gas, i.e. the abundance by mass.
The first item on the right hand side of the equation, $\psi(t)X_i(t)$, represents the mass 
change of the element  \emph{i} caused by the formation of new stars; $R_{i}(t)$ corresponds to 
the material returned through stellar winds or SN explosions in the ISM.
The last two terms take into account the infall of primordial gas $\dot {G}_{i,inf}(t)$ and the 
outflow $\dot {G}_{i,w}(t)$, respectively. 

In particular, the accretion rate of an element \emph{i} follows the equation:
\begin{equation} %\label{infall}
\dot {G}_{i,inf}(t)= {A\,X_{i,inf} e^{-t/ \tau} \over M_{tot}},
\end{equation}
where $X_{i,inf}$ represents the abundance of the element \emph{i} in the infalling gas 
that we assume to be primordial, $A$ is the normalization constant constrained  to reproduce 
the present time total mass (gas plus stars), i.e. $_{tot}$,   and $\tau$ is the accretion timescale expressed in $Gyr$ and defined 
as the characteristic time at which half of the total mass of the system has been assembled.

The rate of gas lost via galactic winds for each specific element \emph{i} is assumed to be 
proportional to the amount of gas present at the time \emph{t} through the free parameter $\lambda_{i}$:
\begin{equation} 
  \dot{G}_{i,w}(t)= \lambda_{i} \psi(t)
\end{equation}

\subsection{Nucleosynthesis prescriptions} 

% No instantaneous recycling approximation is adopted and consequently the stellar lifetimes are 
% properly taken into account. Thanks to this assumption, it is possible to 
We compute in detail the contributions by low mass stars, Type Ia and Type II SNe to the chemical enrichment of the ISM. 
The main physical quantities related to these processes are the stellar yields, representing the 
amount of both newly formed and pre-existing elements ejected into the ISM by stars of all masses 
at their death. \newline
The yields used in this paper are as follows:
\begin{enumerate}
\item for low and intermediate mass stars, LIMS (i.e. with masses $0.8 M_{\odot}\le m \le  8 M_{\odot}$) 
we make use of the yields provided by van den Hoek $\&$ Groenewegen (1997) as a function of metallicity.
\item For massive stars (i.e. with masses larger than $8 M_{\odot}$) and Type Ia SNe, we adopt the 
empirical yields suggested by Fran\c cois et al. (2004), which are substantially based on the 
Woosley \& Weaver (1995) and Iwamoto et al. (1999) yields, respectively, and are tuned to best fit
the abundances in the Solar Neighborhood.
\end{enumerate}

\section{Dust model} \label{dust}

According to Dwek (1998), Calura et al. (2008) and Pipino et al. (2011), the equation describing 
the evolution of the element \emph{i} in the dust is similar to the equation \ref{chem-ISM}.
Let us define $X_{i,dust}(t)$ as the abundance by mass of the element $i$ at the time $t$ locked up 
in the dust; the quantity $G_{dust,i}=X_{dust,i}\cdot G(t)$ represents the normalized mass of 
the element $i$ at the time $t$ in the dust. 
%while $G(t)$ is the ISM fraction at the same time.
By means of the following equation it is possible to understand how a specific chemical element 
evolves in the dust:
\begin{equation} \label{chem-DUST}
\begin{split}
\dot {G}_{i,dust}(t)& = -\psi(t)X_{i,dust}(t) + R_{i,dust}(t)  + (\dot {G}_{i,dust}(t))_{accr} \\ %\,   \nonumber \\
& \quad - (\dot {G}_{i,dust}(t))_{destr}  \quad - (\dot {G}_{i,dust}(t))_{w} \\
\end{split}
\end{equation}
%but, in order to do that, it is important to define the parameters used to describe the processes of 
%formation, accrescion and destruction occurring during the dust evolution.
where, 

\begin{itemize}

 \item $R_{i,dust}(t)$ describes the enrichment rate of the element $i$ where each term of the equation
corresponds to a contribution coming from a different progenitor: 
%  \begin{equation}
%  R_{i,dust}(t) = + \int_{M_{L}}^{M_{B_m}}\psi(t-\tau_m) \delta^{SW}_{i} Q_{mi}(t-\tau_m)\phi(m)dm \\   %\nonumber\\ 
%  & & + A\int_{M_{B_m}}^{M_{B_M}} \phi(m) \cdot[\int_{\mu_{min}}^{0.5}f(\mu)\psi(t-\tau_{m2}) \delta^{Ia}_{i} Q_{mi}(t-\tau_{m2})d\mu]dm \\   %\nonumber \\ 
%  & & + (1-A)\int_{M_{B_m}}^{8 M_{\odot}}\psi(t-\tau_{m})  \delta^{SW}_{i} Q_{mi}(t-\tau_m)\phi(m)dm \\   %\nonumber \\
%  & & + (1-A)\int_{8 M_{\odot}}^{M_{B_M}}\psi(t-\tau_{m})  \delta^{II}_{i} Q_{mi}(t-\tau_m)\phi(m)dm \\   %\nonumber \\
%  & & + \int_{M_{B_M}}^{M_U}\psi(t-\tau_m)  \delta^{II}_{i} Q_{mi}(t-\tau_m) \phi(m)dm \\   %\nonumber\\ 
%  \end{equation}
\begin{align*} 
 R_{i,dust}(t) = &+ \int_{M_{L}}^{M_{B_m}}\psi(t-\tau_m) \delta^{SW}_{i} Q_{mi}(t-\tau_m)\phi(m)dm \\   %\nonumber\\ 
 &+ A\int_{M_{B_m}}^{M_{B_M}} \phi(m) \cdot \\
 & [\int_{\mu_{min}}^{0.5}f(\mu)\psi(t-\tau_{m2}) \delta^{Ia}_{i} Q_{mi}(t-\tau_{m2})d\mu]dm \\   %\nonumber \\ 
 &+ (1-A)\int_{M_{B_m}}^{8 M_{\odot}}\psi(t-\tau_{m})  \delta^{SW}_{i} Q_{mi}(t-\tau_m)\phi(m)dm +\\   %\nonumber \\
 & (1-A)\int_{8 M_{\odot}}^{M_{B_M}}\psi(t-\tau_{m})  \delta^{II}_{i} Q_{mi}(t-\tau_m)\phi(m)dm \\   %\nonumber \\
 &+ \int_{M_{B_M}}^{M_U}\psi(t-\tau_m)  \delta^{II}_{i} Q_{mi}(t-\tau_m) \phi(m)dm    %\nonumber\\ 
\end{align*}

The first term on the right takes into account the enrichment due to stars in the $[M_L-M_{Bm}]$
mass range, where $M_{L}=0.8M_{\odot}$ is the minimum mass contributing, at a given time $t$, to chemical enrichment 
and $M_{B_m}$ is the minimum total binary mass allowed for binary systems 
giving rise to Type Ia SN ($3 M_{\odot}$), while  the mass $M_{BM}=16M_{\odot}$ represents the maximum 
total mass for a binary system to give rise to a SN Ia ($8+8M_{\odot}$) (see Matteucci \& Greggio 1986).
The initial mass function (IMF) used is the Salpeter (1955) one: 
\begin{equation}
\phi (m)\propto m^{-(1+1.35)} 
\end{equation}
The quantity  $Q_{mi}(t-\tau_m)$ (where $\tau_m$ is the lifetime of a star of mass $m$) contains all
the information about stellar nucleosynthesis for elements either produced or destroyed inside 
stars or both (Talbot and Arnett 1973). The second term represents the contribution from binary systems 
that become Type Ia supernovae; $A$ is the unknown fraction of binary stars giving rise to SNIa and is 
fixed by reproducing the observed present time SNIa rate; $t_{m_2}$ indicates the lifetime of the 
secondary star of the binary system and consequently, the explosion timescale is taken into account. 
 The quantity $\mu = M_2/M_B$ is the ratio between the secondary component of the binary system 
(i.e. the less massive one) and the total mass of the system $M_B=M_{Bm} + M_{BM}$
and $f(\mu)$ is the distribution function of this ratio.
The third and fourth terms represent the contribution to dust from single stars in the mass range 
$M_{Bm}-8$ and $8-M_{BM}$ where there are binary systems producing SNIa, single AGB stars and core-collapse SNe.
 \item $(\dot {G}_{i,dust}(t))_{accr} = \frac {{G}_{i,dust}(t)_{accr}} {\tau_{accr}}$ is the rate at which 
 an element \emph{i} is removed from  the ISM gas phase by accretion onto preexisting dust particles in molecular 
 cloud, while $\tau_{accr}$ is the timescale for the accretion process that will be described in section \ref{accr}. 
 \item $(\dot {G}_{i,dust}(t))_{destr} $ rapresents the amount of the element 
 \emph{i} returned to the ISM gas phase, owing to the dust destroyed by SNe, on the timescale of destruction $\tau_{destr}$ (see \ref{destr}).
 \item $(\dot {G}_{i,dust}(t))_{w}$ takes into account  possible ejection of dust into the inter galactic medium (IGM) 
 by means of galactic winds   triggered by SNe explosions. The wind starts when the thermal energy of gas equates the binding 
 energy of gas. In our calculations, we assume a differential wind in which the metals are preferentially lost (see Bradamante et al. 1998).
\item the quantities $\delta_i$ represent the dust condensation efficiencies as defined by Dwek (1998) and are relative to the low and 
intermediate mass stars which eject dust through stellar winds (SW), the Type II SNe (II) and the Type Ia SNe (Ia). For the model of the ellipticals we assume that after the 
onset of the galactic wind the star formation is quenched and the galaxy starts evolving passively, whereas for spirals and irregulars 
the star formation continues even after the wind.

\end{itemize}

%finisci di spiegare i parametri

\subsection{Supernovae rates} \label{SNrate}
The evolution of dust is strongly dependent on the SN rates (SNRs). In our computation we assume, as contributors to the total dust amount 
of a galaxy, both core collapse (CC) and Type Ia supernovae. 
CCSNe are divided in Type II and Ib/c SNe; in particular Type II SNe, originate from single stars with  mass larger than $M_w \sim 8M_{\odot}$ 
while SNe Ib/c come from  Wolf-Rayet (WR) stars, namely stars which have lost most of their H and He envelope and with masses larger than $M_{WR} = [25 \div 40]M_{\odot}$, 
whose value depends on the initial stellar metallicity. 
% In fact, the mass loss in massive stars ($M \ge 10M_{\odot}$) increases with the initial metallicity in a way 
% that $M_{WR}$ decreases with increasing metallicity
It has been suggested that they could be both single WR stars and  massive stars in binary systems in the mass range 14.8-45$M_{\odot}$, (Yoon et al. 2010; Grieco et al. 2012).
Owing to the larger mass of the progenitors, CCSNe explode on short timescales: from few million to several tenths of million years.

Type Ia SNe are believed to originate from low and intermediate mass stars, in particular from C-O white dwarfs in binary systems.
%,ending their lives on timescales ranging from $\sim 0.03$ up to an Hubble time or more. 
In particular, to describe Type Ia supernova progenitors, we assume the single degenerate (SD) scenario by Whelan \& Iben (1973). 
In this scenario, a C-O white dwarf accretes mass from a non-degenerate companion until it reaches the Chandrasekhar mass ($∼ 1.4M_{\odot}$) and explodes via 
C-deﬂagration, leaving no remnant.
Type Ia SNe end their lives on timescales ranging from $\sim 0.03 \, Gyr$ up to an Hubble time or more and consequently the SNIa rate is related to the past star formation 
history (SFH) of a galaxy, whereas the CCSNe reflect the birth rate of massive stars, i.e. the galactic star formation rate (SFR). 
For this reason, different types of galaxies have different SNRs.

The total SNR used in our models is the sum of the three individual rates, that may be approximated as :

\begin{equation} \label{SNII_eq}
  R_{SNII}(t)= \int_{8}^{M_{WR}} \psi(t-\tau_m)\phi(m)dm   
\end{equation}
\smallskip
\begin{equation} \label{SNIbc_eq}
 \begin{split}
  R_{SNIb/c}(t) &= \int_{M_{WR}}^{100} \psi(t-\tau_m) \phi(m)dm + \\
             &\quad + F \int_{14.8}^{45} \psi(t-\tau_m) \phi(m)dm \\
 %            & \sim \psi(t) ( \int_{M_{WR}}^{100} \phi(M)dM + F \int_{14.8}^{45} \phi(M)dM )
 \end{split}
 \end{equation}
\smallskip
\begin{equation} \label{SNIa_eq}
  R_{SNIa}(t)= A \int_{M_{Bm}}^{M_{BM}} \phi(M_{B}) [\int_{\mu_{m}}^{0.5} f(\mu)
\psi(t-\tau_{m_{2}})d\mu]dM_{B}
\end{equation}

The SN Ib/c rate is composed by a first term evaluating the single progenitor channel and a second integral that takes into account
the contribution from WRs in binary system. The factor F represents the fraction of massive binary stars producing 
Type Ib/c SNe. This parameter is chosen to be equal to 0.15, (see Calura \& Matteucci, 2006). 
This value is motivated by the facts that half of the massive stars are possibly in binary systems, 
and the fraction of massive stars in close binary system is $\sim 30\%$, i.e. similar to the close binary frequency 
predicted for low mass systems (Jeffries \&  Maxted, 2005).  Therefore, the estimated value for this parameter is  $F \sim 0.5 \cdot 0.3 \sim 0.15$.

\subsection{Dust production}  \label{prod}
There are still many uncertainties in the yield of dust from SNe and AGB stars. 
The general definition of dust grains divide the elements that suffer depletion into dust (i.e. refractory elements)
into silicate dust, composed by O, Mg, Si, S, Ca, Fe and Ni, and into carbon dust, composed only by C.
 
%metti Hirashita 2012 grain size distribution (SNe source)
%metti zhukovska2007 e piovan per differenti valori di delta

Following Dwek (1998) the dust sources considered in our work are:
\begin{itemize}
 \item \emph {Low and Intermediate mass stars (LIMS)}:
 in these stars, dust is produced during the AGB phase. 
 We assume that dust formation depends mainly on the composition of the stellar envelopes. 
If $X_{O}$ and $X_{C}$ represent the O and C mass fractions in the stellar envelopes, respectively, we assume that 
stars with $X_{O}/X_{C}$ $>$ 1 are producers of silicate dust, i.e. dust particles composed by O, Mg, Si, S, Ca, Ni, Fe. 
On the other hand, C rich stars, characterized by $X_{O}/X_{C}$ $<$ 1, are producers of carbonaceous solids, 
i.e. carbon dust (Draine 1990). Being $M_{i, ej}(m)$ and $M_{i, dust}(m)$ the total ejected mass and the dust mass 
formed by the stars as functions of the initial mass $m$ for the element $i$, 
respectively, we assume that for stars with $X_{O}/X_{C}$ $<$ 1: 
\\
\\

\begin{math}
M_{C, dust}(m) =  \delta^{SW}_{C} \cdot [M_{C, ej}(m)-0.75 M_{O, ej}(m)]\\
\end{math}
\\
with $\delta^{SW}_{C}=1$ and 
\\
\\
\begin{math}
M_{i, dust}(m) =0, \\
\end{math}
\\
for all the other elements. For stars with $X_{O}/X_{C}$ $>$ 1 in the envelope, we assume 
\\
\\
\begin{math}
M_{C, dust}(m) =0 \\
\end{math}
\\
\\
\begin{math}
M_{i, dust}(m) = \delta^{SW}_{i} M_{i,ej}(m) \\
\end{math}
\\
\\
with $\delta^{SW}_{i}=1$ for Mg, Si, S, Ca, Ni, Fe and \\
\\
\\
\begin{math}
M_{O, dust}(m)=16 \sum_{i} \delta^{SW}_{i} M_{i,ej}(m)/\mu_{i}   \\
\end{math}
\\
\\
with $\mu_{i}$ being the mass of the $i$ element in atomic mass units. 
%\\
 \item \emph {Supernovae (SNe)}:
 SNe are potentially the most important source of interstellar dust. 
 Core collapse supernovae (see section \ref{SNrate} and in particular Type II supernovae (SNe II)
 are considered to be one of the main grain production sources in the Universe (e.g. Nozawa et al. 2003). 
 Assuming that all the refractory elements precipitate with a mean efficiency of $50\%$, a typical 
 $25 M_{\odot}$ SN can produce about $0.5M_{\odot}$ of dust (Woosley \& Weaver 1995; Nomoto et al. 2006). %FAI IL CONTO CON PRECISIONE,dwek07!!!
 % da qui.... referenze da szalai et al 2011 
 The first evidence for dust condensation was observed in SN 1987A (Danziger et al. 1989, Lucy et al. 1989);
 formation of dust was then observed in several CCSNe, i.e. SN 2003gd (Meikle et al. 2007), 2004et 
 (Kotak et al. 2009), 2004dj (Szalai et al. 2011), 2007od (Andrews et al. 2010) and Type Ib/c SN 2006jc (Nozawa et al. 2008; Mattila et
al. 2008; Tominaga et al. 2008 and Sakon et al. 2009), with an estimated mass of recently formed dust between
$10^{-5} - 10^{-3} M_{\odot}$. 

From a theoretical point of view, the amount of dust formed after a CCSN explosion was evaluated by the models of Kozasa et al (1989), 
Todini \& Ferrara (2001) and Nozawa et al. (2003) in the range $0.1-1 M_{\odot}$. This prediction was confirmed
by numerical models of Bianchi \& Schneider (2007), Kozasa et al. (2009) and Silvia et al. (2010). 
The discrepancy between observations and theories is still unclear and also the study of SN remnants could not solve the
question of the amount of dust produced by SNe. Using far-infrared and sub-millimeter data, many groups estimated the
amount of dust in Cas A (Dunne et al. 2003; Krause et al. 2004; Rho et al. 2008). Their results varied between $0.02$ and $2 M_{\odot}$, 
while the values for Kepler SNR showed differencies from  $0.1 - 3\, M_{\odot}$ (Morgan et al. 2003) down to
$5 \cdot 10^{-4} M_{\odot}$ (Blair et al. 2007). 
Recently, Matsuura et al. (2011) report far-IR and sub-mm observations of SN1987A in the Large Magellanic Cloud. The
observations revealed the presence of cold dust grains with a temperature of $T_d \sim 17-23\,K$.
The intensity and spectral energy distribution of the emission suggests a dust mass of $0.4-0.7\,M_{\odot}$.
% ....a qui
 
It has been supposed that not only CCSNe but also SNeIa can be possible
producers of dust grains, especially Fe grains (Tielens 1998).

Following Calura et al. (2008), we assume the same prescriptions for both
SNe types: 
 
% Here , we have assumed, in agreement with Dwek (1998) and Calura et al. (2008), that not only CCSNe but also SNeIa can be possible producers of dust grains, 
% especially Fe grains (Tielens 1998). Recently, Gomez et al. (2012) have observed dust in historical Type Ia SNe and concluded that there is a 
% lack of cold dust grains in Type Ia remnants argues against a substantial production of iron-rich dust grains. 
% Although they cannot completely rule out a small mass of freshly formed supernova dust, the Herschel observations seem to confirm 
% that significantly less dust forms in the ejecta of Type Ia supernovae than in the remnants of core-collapse explosions. 

\begin{math}
M_{dust, C}(m) = \delta^{Ia,II}_{C}[M_{C,ej}(m)]\\
\end{math}

with  $\delta^{Ia,II}_{C}=0.7$; 

\begin{math}
M_{dust, i}(m) = \delta^{Ia,II}_{i}M_{i,ej}(m)\\
\end{math}

with $\delta^{Ia,II}_{i}=0.8 $ for Fe, Ni \\

$\delta^{Ia,II}_{i}=0.7 $ for  Mg, Si, Ca  and \\

with $\delta^{Ia,II}_{i}=0.5 $ for S; \\

\begin{math}
M_{dust, O}(m)=16 \sum_{i} \delta^{Ia,II}_{i} M_{i,ej}(m)/\mu_{i}  \\
\end{math}

%Respect Dwek (1998), we use a different prescription for silicon and sulfur.
%  motivazione, magari prova ad usare piovan e zhukovska2007

There are still many uncertainties on the amount of dust produced by Type
Ia SNe. However, Gomez et al. (2012), on the basis of Herschel PACS and
SPIRE photometry at $70-500 \mu m$ have reported the existence of dust in
two Type Ia SN remnants: Kepler and Tycho. In particular, they detected a
warm dust component in Kepler's remnant with temperature $T_d=82_{-6}^{+4}\,K$
and mass $M_d \sim 3.1_{-0.6}^{+0.8} \cdot 10^{-3} M_{\odot}$; similarly for the
Tycho's remnant, they detected warm dust at $T_d=90_{-7}^{+5}\,K$ with mass
 $M_d \sim 8.6_{-1.8}^{+2.3} \cdot 10^{-3} M_{\odot}$.

 \end{itemize}

\subsection{Dust accretion} \label{accr}

Dust grains can grow by accretion of metals present in the ISM onto preexisting grain cores within dense molecular clouds (Dwek 1998, Inoue 2003).
There are more observations indicating the existence of large, micrometre-sized dust grains in dense molecular clouds 
and suggesting a need for significant dust growth in the early Universe (Pagani et al. 2010, Michalowski et al. 2010). 
Moreover, dust growth appears to dominate over dust destruction also in the local, present-day Universe (Hirashita 1999; Inoue 2003; 
Hirashita \& Kuo 2011);  
other evidence for dust accretion comes from the observed infrared emission of cold molecular clouds (Flagey et al. 2006).

Following Dwek (1998), we consider that the key parameter in dust accretion is the dust accretion timescale that,
for a given element $i$, can be expressed as an increasing function of the dust mass: 

\begin{equation}
\tau_{accr}=\tau_{0,i}/(1 - f_i) 
\label{accr_t}
\end{equation} 

where 
\begin{equation}
f_i=\frac{G_{dust,i}}{G_{i}}
\end{equation} 
 
For the timescale $\tau_{0,i}$, typical values span from $\sim 5 \times 10^{7}$ yr, of the order of 
the lifetime of a typical molecular cloud, up to $\sim 2 \times 10^{8}$ yr. 
In this paper, we assume that the timescale $\tau_{0,i}$ is constant for all elements, 
with a value of $5 \times 10^{7}$ yr.\\

%visto che hai provato anche a cambiarli magari sarebbe utile mettere qualche riga sui risultati.

\subsection{Dust destruction} \label{destr}

The main mechanism for dust destruction is by sputtering in the high interstellar shocks driven by SNe (McKee 1989; Jones et al 1994). 
Following Dwek et al. (2007), the dust-destruction timescale is independent of the dust mass and equal to:
\begin{equation} 
  \tau_{destr,i}= \frac {\sigma_{gas}}{<m_{ISM}> R_{SN}}
\end{equation}
where $\sigma_{gas}$ is the gas mass density, $<m_{ISM}>$ is the effective ISM mass that is completely cleared 
of dust by each SN event, and $R_{SN}$ is the sum of core collapse and type Ia SN rates.
Dwek et al. (2007) consider the grain destruction efficiency, corresponding to the range of uncertainty in the lifetime of 
the interstellar dust grains, as an unknown, and adopt $<m_{ISM}>$ as a free parameter of the model. 
Regarding the Milky Way like galaxies, for values of $n_0 \sim 0.1 - 1.0 cm^{-3} $, corresponding to the average density of
the Galactic ISM, they get  $<m_{ISM}>$ = 1100 - 1300 $M_{\odot}$ for an equal mix of silicate and
graphite grains.

 \section{The host sample} \label{sample} 

In order to constrain the SF history of the HGs we have chosen from the literature
the following HGs with detailed observed abundance data: GRB 081008 (D'Elia
et al. 2011), 120327A (D'Elia et al. 2014) and 120815 (Kruhler et al. 2013). In
Table \ref{grb_data} are shown the observational data for each HGs studied in our analysis.
Concerning the [O/Fe] and [Al/Fe] abundances of GRB 081008 (first column),
these values are considered as lower limits because of the lines saturation (see
D'Elia et al. 2011 for more details). The same problem is in the C, O and Al
abundances in fact in the D'Elia et al. (2014) paper, the authors warn the reader
explaining that despite the highly saturated, CII $\lambda$1334, OI $\lambda$1039, OI $\lambda$1302 and
AlII $\lambda$1670 transitions have not been excluded from the analysis, because they
are the only lines describing CII, OI and AlII.

\begin{table}
\begin{center}
\begin{tabular}{|c|c|c|c|}
 \hline
Quantity              &    GRB 081008        &    GRB 120327A        &    GRB 120815       \\
\hline
redshift z            &        1.968         &         2.81          &        2.36         \\
\hline
$M_* \, [M_{\odot}]$  &        -             &                       &   $\leq10^{10}$     \\    
\hline
[Fe/H]                &   -1.19 $\pm$ 0.11   &   -1.73 $\pm$ 0.10     &   -2.19 $\pm$ 0.12  \\
\hline
[Mg/Fe]               &        -             &   0.46 $\pm$ 0.05     &         -           \\
\hline
%[O/Fe]                &  $>$ -1.49 $\pm$ 0.09&   -0.25 $\pm$ 0.1(S)     &         -           \\
%\hline
[Si/Fe]               &   0.32 $\pm$ 0.07    &   0.57 $\pm$ 0.05     &   $\ge$1.04 $\pm$ 0.17   \\
\hline
[S/Fe]                &        -             &   0.34 $\pm$ 0.04     &   $\le$1.31 $\pm$ 0.26   \\
\hline
[N/Fe]                &        -             &   0.28 $\pm$ 0.08     &         -           \\  
\hline
%[C/Fe]                &        -             &   1.94 $\pm$ 0.16(S)     &         -           \\     
%\hline
[Ni/Fe]               &   -0.4 $\pm$ 0.09    &   0.11 $\pm$ 0.05     &   0.18 $\pm$ 0.09    \\
\hline
[Zn/Fe]               &   0.67 $\pm$ 0.07    &   0.56 $\pm$0.07      &   1.12 $\pm$ 0.10      \\
%\hline
%[Ca/Fe]               &        -             &   0.39 $\pm$ 0.28     &         -           \\      
%\hline
%[Al/Fe]               &$>$ -0.67 $\pm$ 0.15  &   1.53 $\pm$ 0.18     &         -           \\  
%\hline
%[Cr/Fe]               &   0.27 $\pm$ 0.15    &   0.26 $\pm$ 0.14     &   0.28 $\pm$ 0.16   \\
%\hline
%[Mn/Fe]               &        -             &         -             &   0.0 $\pm$ 0.15    \\
\hline
\end{tabular}
\end{center}
\caption{Observational data collected for GRB 081008 (D'Elia et al. 2011), 120327A (D'Elia et al. 2014) and 120815 
(Kruhler et al. 2013). Note that all the abundances have been  normalized to the same solar abundances, namely those of Asplund et al. (2009).}
\label{grb_data}
\end{table}

\section[]{Model results} \label{res}

\begin{figure*} 
\begin{center}
\includegraphics[scale=0.35]{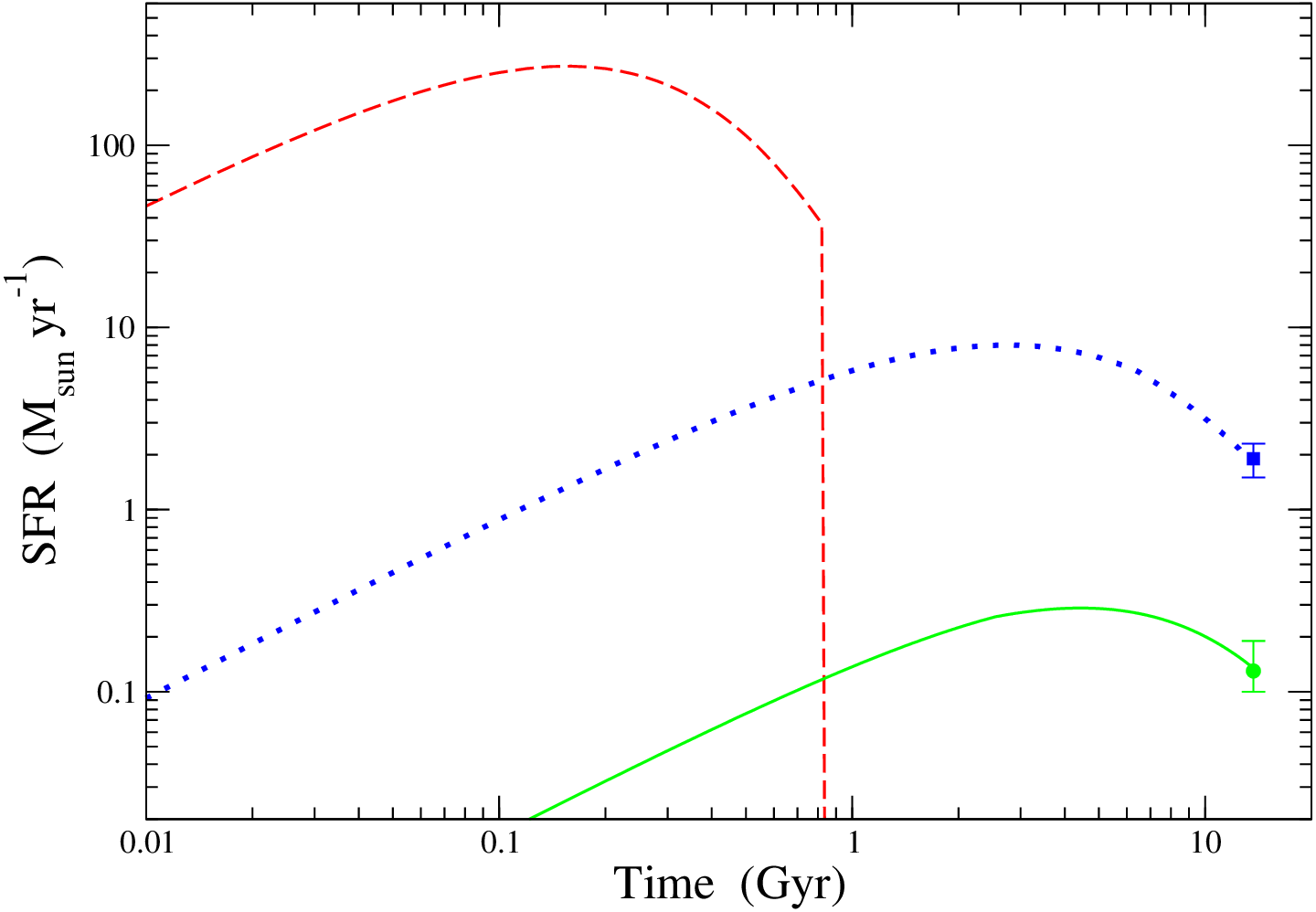}      
\caption{Predicted SFRs for galaxies of different morphological type.
The three curves are obtained by varying the efficiency of star formation:
$\nu_{E00}=10 Gyr^{-1}$ for the elliptical, $\nu_{S00}=1 Gyr^{-1}$ for the spiral, and $\nu_{C00}=0.1 Gyr^{-1}$ for the irregular galaxy.
%The infall masses for each type of galaxies are: $10^{11} M_{\odot}$ for ellipticals (dashed line), $5\cdot 10^{10}M_{\odot}$ for spirals (dotted line) 
%and  $5\cdot 10^{9}M_{\odot}$ for irregulars (solid line).
The two points with error bars refer to measured average star formation rate at the present time in spirals 
(square,  Chomiuk \& Povich, 2011)) and irregulars (circle, Harris \& Zaritsky, 2009)}.
\label{SFR}
\end{center}
\end{figure*}

We run several models for galaxies of different morphological type. In particular,
in Table \ref{models-param} are summarized the parameter sets of each specific model with the
dust prescription, as described in Section \ref{dust}; the two ellipticals models (E00, E01)
describe the evolution of massive spheroids, the model S00 represents a typical spiral galaxy Milky Way-like and model C00
a typical irregular model. In the first column is the infall mass of
the system ($M_{tot}$), i.e. the final total assembled mass if nothing is lost; in the
second column is the efficiency of star formation (SFE) which characterizes the
different galaxy types, and finally, in the third column, the dust yields 
($\delta_i$, $i = C,S,Si,Mg,Ca,Fe,Ni$) for Type Ia and core collapse SNe (see section \ref{prod})
are shown. Note that for elliptical galaxies we show two models with different infall
mass and star formation efficiency (SFE). One of the most common assumptions
used in this work is the increase of the SFE with the galactic mass among different
morphological galaxy types. This assumption was first suggested by Matteucci
(1994) to explain the increase of the [Mg/Fe] ratio with galactic mass in ellipticals
(downsizing in star formation); in fact, different SFEs
produce different SFRs as shown in Figure \ref{SFR} and different abundances and abundance ratios. In Figure \ref{SFR} are shown  
the SFRs of model E00, S00 and C00 as  functions of the galactic time. 
Note that in model E00 the star formation stops after the onset of the galactic wind, at $t_W \sim 0.5$ Gyr. 
These three galaxy models will represent typical galaxies for each morphological since now on, although for 
ellipticals we adopt also model E01 as representative of a very massive spheroid. 
We recall here  that the models presented are very similar to those described in Calura et al. (2008). 
In that paper, as a sanity check for the models, it was shown that the spiral model, when applied to the solar region,  
could reproduce the observed depletion pattern in the local interstellar cloud (Kimura et al. 2003). 
So we adress the interested reader to the Calura et al.(2008) paper for an extensive discussion on this point. 
Our present model for a spiral represents an average spiral galaxy but if applied to the solar neighbourhood it 
predicts exactly the same values of Figure 3 of Calura et al. (2008). By the way,  the more recent dust to metal ratio 
derived by De Cia et al. (2013) for the solar vicinity is also in good agreement with our results.

\begin{table}
\begin{center}
\begin{tabular}{|c|c|c|c|}
 \hline
             & $M_{tot}\, [M_{\odot}]$ & $SFE$ \, $[\mbox{Gyr}^{-1}]$   &          SN Dust Yields           \\       
%  \hline
% E00          &       $10^{11}$         &           $10$               &         $\delta_{C}=0.5$           \\
%  \cline{1-3}
% S00          &       $10^{10}$         &           $1$                &         $\delta_{i}=0.8$,           \\
%  \cline{1-3}
% C00          &       $10^{9}$          &          $0.1$               &        i=S,Si,Mg,Ca,Fe,Ni           \\
% \hline  
E00          &       $10^{11}$            &           $10$               &                                 \\
 \cline{1-3}
E01          &       $5 \cdot 10^{11}$         &           $20$               &  $\delta_{i}=0.7$, i=C,Mg,Ca,Si  \\
 \cline{1-3}
S00          &       $5 \cdot 10^{10}$         &           $1$                &  $\delta_{i}=0.5$, i=Fe,Ni       \\
 \cline{1-3}
C00          &       $5 \cdot 10^{9}$          &          $0.1$               &     $\delta_{S}=0.5$             \\
\hline  
\end{tabular}
\end{center}
\caption{Parameters adopted for describing our models: $M_{tot}$ is the infall mass, namely the final total mass that should be assembled if nothing is lost,
 $SFE$ is the star formation efficiency and $\delta$ is the SN (Ia + II) dust yield of the considered elements (C,S,Si,Mg,Ca,Fe,Ni).}
\label{models-param}
\end{table}  

\subsection[]{The dust rates in galaxies}

\begin{figure*} 
\begin{center}
\includegraphics[scale=0.35]{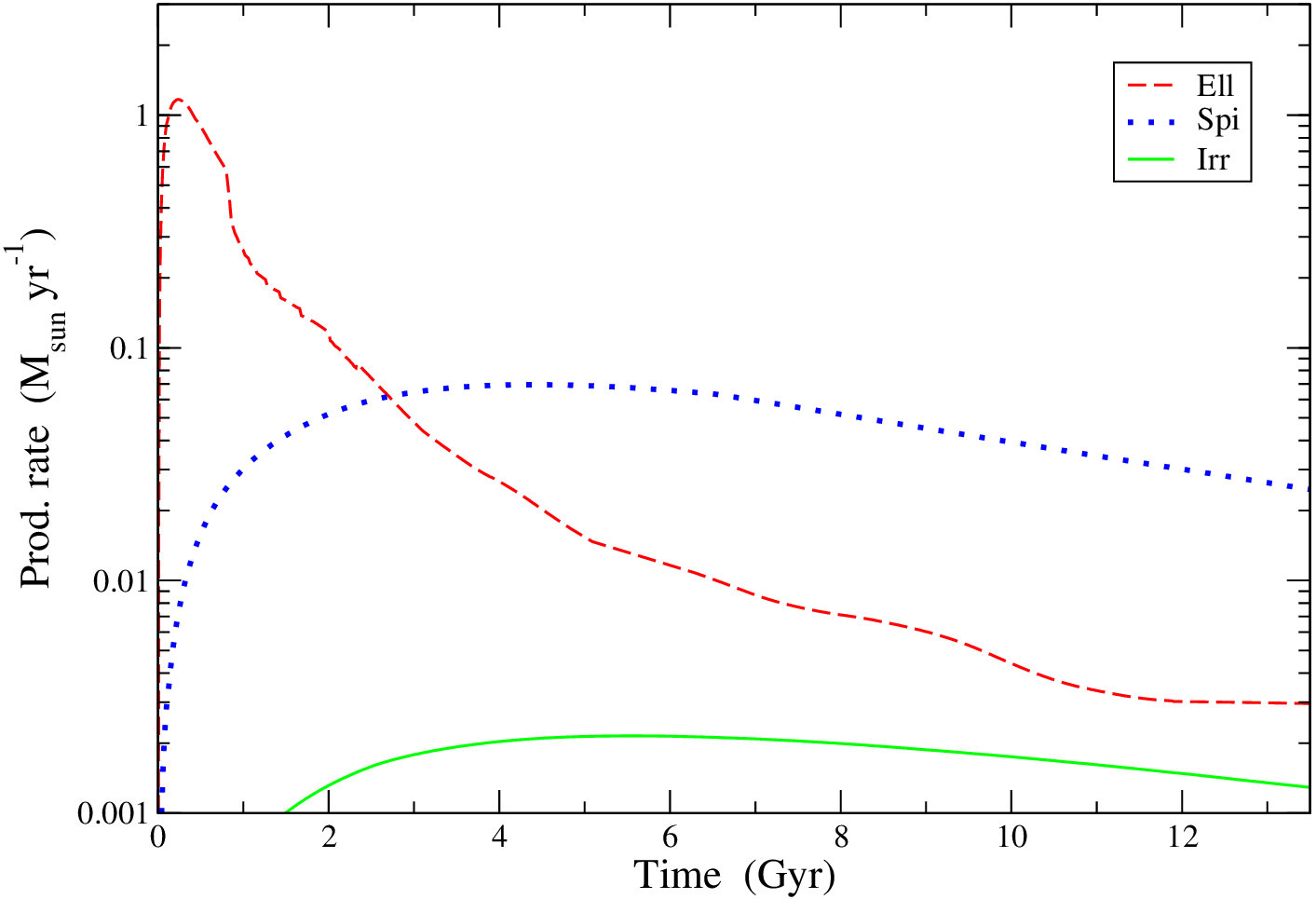}     
\caption{ Rates of dust production by stars in galaxies of different morphological type: the black continuous line 
refers to ellipticals; the lines are indicated with the same symbols as in Fig. 1.}
\label{ProdRate}
\end{center}
\end{figure*}

\begin{figure*} 
\begin{center}
\includegraphics[scale=0.35]{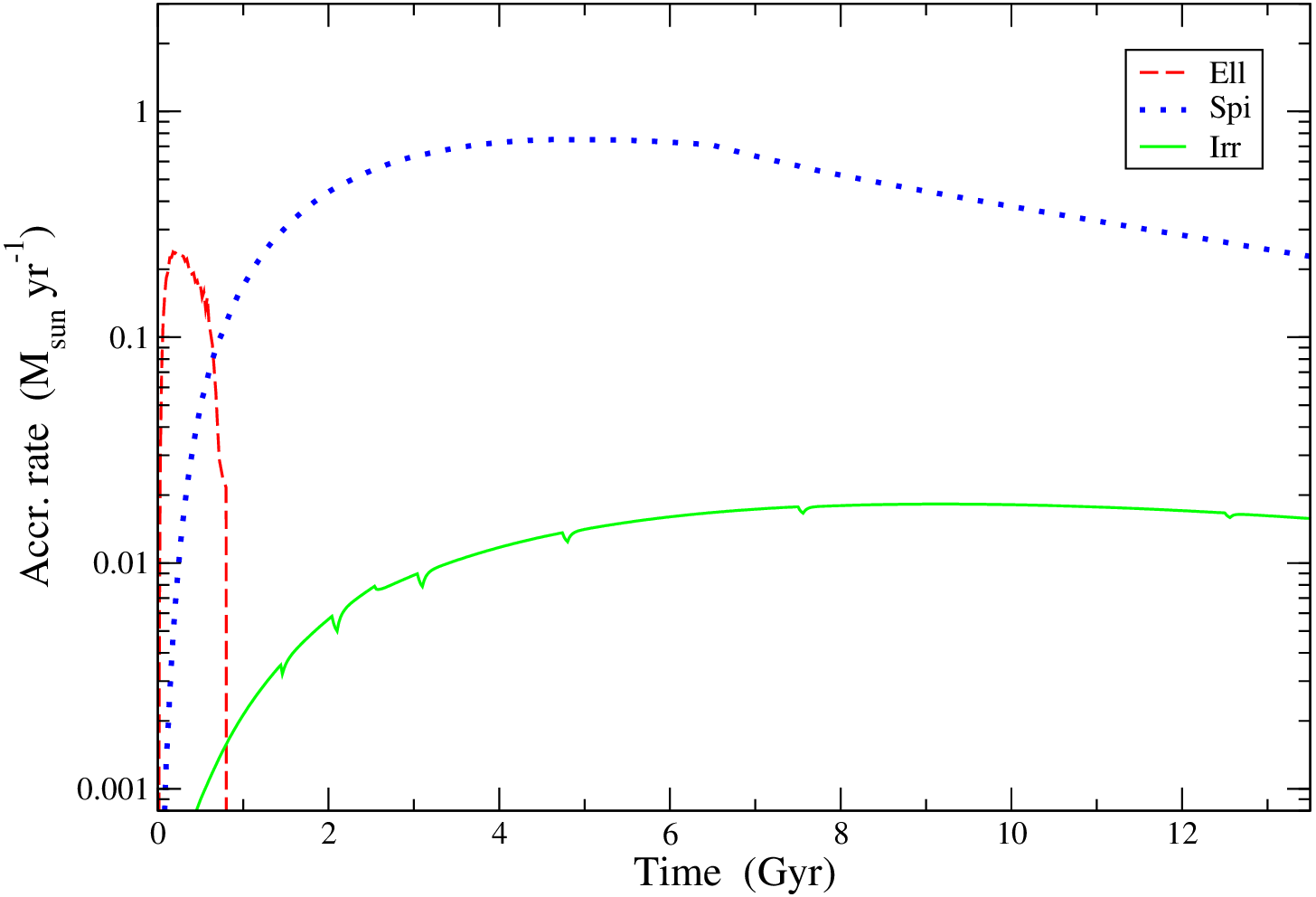}     
\caption{ Rates of dust accretion in galaxies of different morphological type: the black continuous line refers to ellipticals;  
the lines are indicated with the same symbols as in Fig. 1.}
\label{AccrRate}
\end{center}
\end{figure*}

\begin{figure*} 
\begin{center}
\includegraphics[scale=0.35]{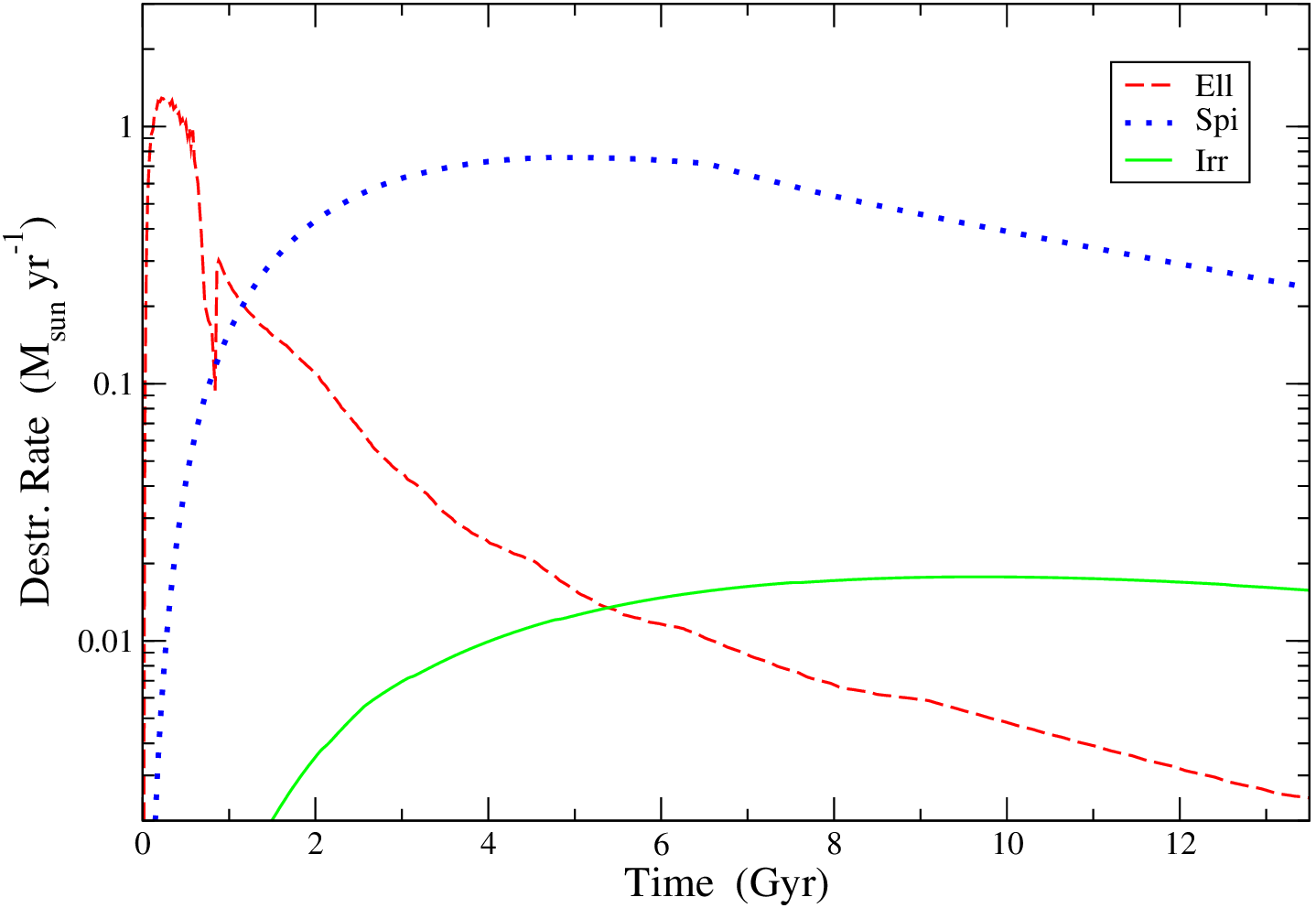}     
\caption{ Rates of dust destruction in galaxies of different morphological type: the black continuous line refers to ellipticals; 
the lines are indicated with the same symbols as in Fig. 1.}
\label{DestrRate}
\end{center}
\end{figure*}

As the dust content in the Universe is the result of different and competitive processes, 
we study the individual contribution of accretion, destruction
and production in each morphological type of galaxy using the model E00 with
$M_{tot} = 10^{11} M_{\odot}$ and SFE=$10\,Gyr^{-1}$ for an elliptical, the model S00 with
$M_{tot} = 5 \cdot 10^{10} M_{\odot}$ and SFE=$1\,Gyr^{-1}$ for a spiral and finally the model 
C00 with $M_{tot} = 5 \cdot 10^{9} M_{\odot}$ and SFE=$0.1\,Gyr^{-1}$ for an irregular galaxy. We focus
on the effects that different star formation histories have on the evolution of the
chemical abundances in the ISM and how these abundances are affected by dust
in galaxies of different morphological type as well as on the different competitive
rates involving the dust. In particular, understanding how the dust rates evolve
in time is important in the computation of the global amount of dust and the
cosmic rate inferred from our models. The main assumption concerning the dust
rate is on the dust yields describing the fractions of each element followed in our
chemical evolution code, which is condensed into dust and restored into the ISM
by low and intermediate mass stars, Type Ia and II SNe. 
The prescription used to 
infer the dust production rate are summarized in Table \ref{models-param}. 
The rate of dust production in units
of $M_{\odot} yr^{-1}$  is shown in  \ref{ProdRate} and describes the amount of dust produced during the whole galactic
evolution for the elliptical , spiral and irregular
galaxy model. In Figures \ref{AccrRate} and \ref{DestrRate} are shown the
rates of accretion and destruction, respectively, in unit of $M_{\odot} yr^{-1}$ as functions
of time for the same models shown in Figure \ref{ProdRate} and with the same color set.

As one can see from Figures \ref{ProdRate}, \ref{AccrRate} and \ref{DestrRate}, the rate of dust production in 
ellipticals is maximum at the beginning of their evolution 
but also the destruction rate is maximum since they both depend on the SFR, which is assumed to be very high in the first phases 
of the evolution of these objects (see Fig. 1). The dust accretion rate instead is quite high in spirals when the SFR reaches a maximum,
whereas accretion is active in ellipticals only during the starburst and it stops after the occurrence of a wind which devoids 
the galaxy of gas and dust. On the other hand, the rate of destruction of dust in ellipticals continue even after the star formation 
has stopped, since it depends on SN explosions and Type Ia SNe are active until the present time even in absence of star formation. 
These different behaviours are responsible for the shape of the cosmic effective dust production rate that will be described in the next Section.

\subsection{Abundance ratios}

In order to understand the different behaviour of the chemical evolution models developed here, we compare our model results with the abundances measured 
in the hosts of the three GRBs considered. Our main goal was to derive the history of star formation and therefore the nature of the hosts comparing the 
predictions of [X/Fe] vs.[Fe/H] for several chemical elements (C, N, O, Mg, S, Si, Ni and Zn). 
Among these species, N and Zn are not affected by dust while all the others, together with iron, contribute to the total amount of dust.

\begin{figure*} 
\begin{center}
\includegraphics[scale=0.5]{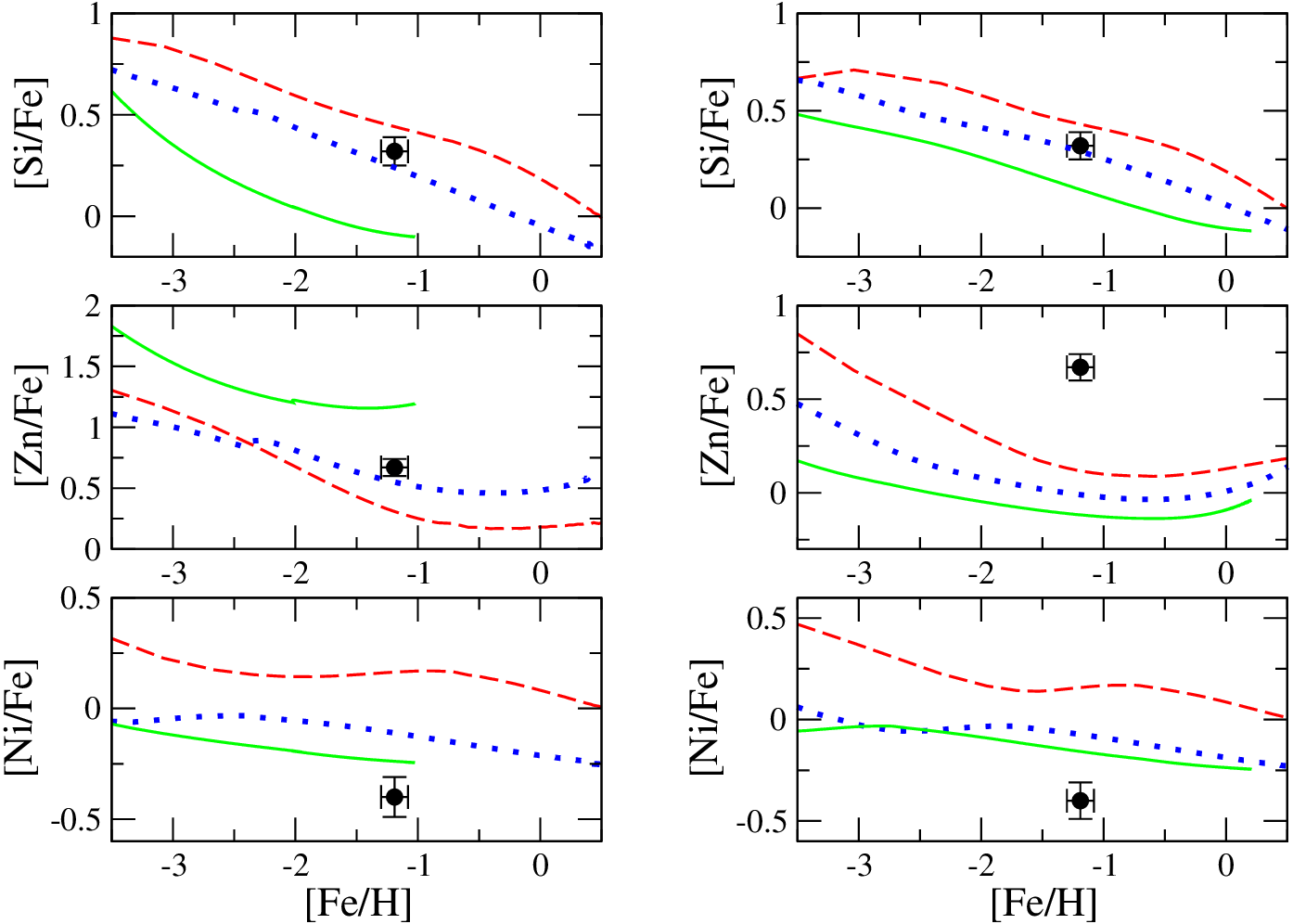}      
\caption{Observed abundance ratios vs. metallicity for the host galaxy of GRB 081008 provided by D'elia et al. (2008). The left column show the 
models predictions including the dust treatment while the right column the same models without dust evolution.
The red-dashed (E00), blue-dotted (S00) and green-solid line (C00), are the predictions computed by means of the chemical evolution models for 
an elliptical, spiral and dwarf irregular galaxy, respectively. The star formation efficiency assumed in these models are
$\nu_{E00}=10 Gyr^{-1}$, $\nu_{S00}=1 Gyr^{-1}$, and $\nu_{C00}=0.1 Gyr^{-1}$. Data relative to O are not reliable and they are not used to constrain the models.}
\label{081008}
\end{center}
\end{figure*}

In Figure \ref{081008}, we show the results for the host of GRB 081008 at  $z=1.968$ studied by D'Elia et al. (2011). 
In order to assess the importance of dust in this kind of analysis, in Figure \ref{081008} (only), are shown the predictions for
$\alpha$ -elements, Zn and Ni relative to Fe both in the models with dust (left column) and  without dust (right column). 
It is worth noting that while the ratio [Si/Fe] does not change in the two cases due to the fact that both elements are depleted in dust, 
for other ratios, such as [Zn/Fe] the situation is different. For example, the curve relative to the irregular galaxy is much lower in the case 
without dust depletion. The reason for that resides in the fact that Zn is not refractory, whereas Fe is. In particular, in the irregular model 
the amount of Fe in dust is large due to the less efficient dust destruction process relative to the other galaxy types. 
This is because the destruction rate depends on the SFR which is quite low in irregulars.Therefore, this implies that at any given metallicity the 
[Zn/Fe] ratio increases relative to the case with no dust depletion.
On the other hand, in the spiral and elliptical models, the [Zn/Fe] is not much altered since they are characterised by a dust destruction rate which
is at least comparable to the accretion rate. In the case without dust the models are allabove the data and this trend suggests that the subtraction 
of Fe by dust is necessary to explain the data. 
 Looking at all the abundance ratios together in Figure \ref{081008},  we cannot derive a firm conclusion on the nature of the host of this GRB: 
in fact, while the [Ni/Fe] ratio is inconsistent with all curves, the [Si/Fe] and [Zn/Fe] seem to indicate a spiral galaxy. 

\begin{figure*} 
\begin{center}
\includegraphics[scale=0.5]{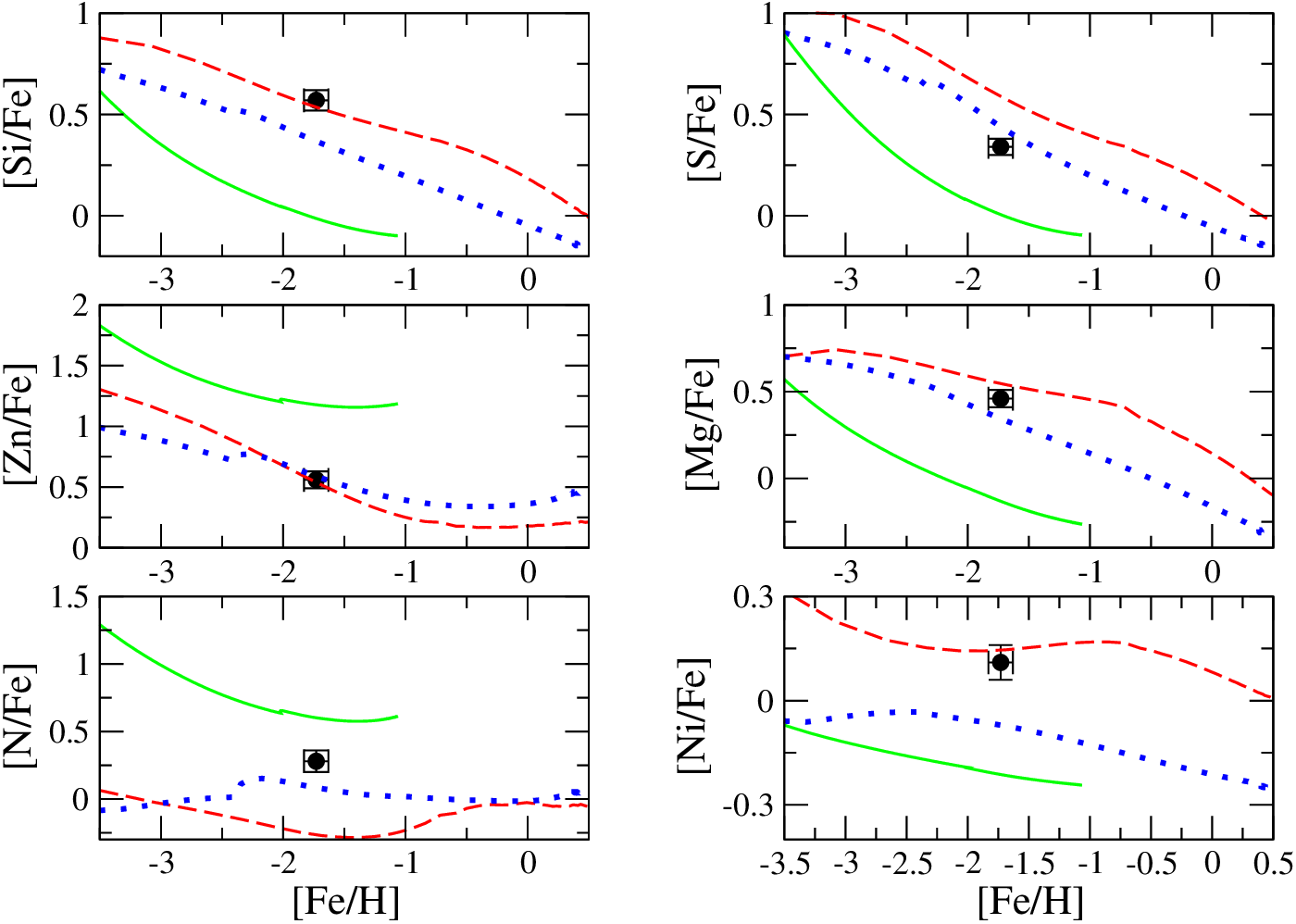}      
\caption{Observed abundance ratios vs. metallicity for the host galaxy of GRB 120327A provided by D'Elia et al. (2014).
The red-dashed (E00), blue-dotted (S00) and green-solid line (C00), are the predictions computed by means of the chemical evolution models for 
an elliptical, spiral and dwarf irregular galaxy, respectively. The star formation efficiency assumed in these models are
$\nu_{E00}=10 Gyr^{-1}$, $\nu_{S00}=1 Gyr^{-1}$, and $\nu_{C00}=0.1 Gyr^{-1}$. Data relative to O and C are not reliable and they are not used to constrain the models.}
\label{120327A}
\end{center}
\end{figure*}

In Figure \ref{120327A} are shown the predictions and the data for the host of GRB 120327A at $z=2.81$ studied by D'Elia et al. (2014). 
In this case we show only the models with dust for several elements. 
Concerning N and Zn it is possible to see an inversion in the models: the irregular galaxy model (C00) predicts larger N and Zn abundances with respect 
to the Fe than the other two models. 
To understand the reason of this behaviour, we need to remember that N and Zn are not affected by dust while Fe is strongly depleted. 
Indeed, the higher is the star formation efficiency of a galaxy, the higher is the dust grain destruction. Therefore, the model with lower SFE 
(green solid line) predicts higher abundances in these two particular cases. 
The trends of [Si,S,Zn/Fe] vs. [Fe/H]  seem to suggest that the host galaxy is a system with high mass and SFR, namely either a spiral or a 
proto-spheroid. Concerning the [N/Fe] ratio it should be said that the N yields are still uncertain and in particular the secondary/primary 
nature of this element. In particular, here  N is considered secondary from massive stars, while there are suggestions that N could be produced 
in a primary fashion in massive stars but only at very low metallicity (see Chiappini et al. 2006). If this is true,  it could change the shapes of the models.

\begin{figure*} 
\begin{center}
\includegraphics[scale=0.5]{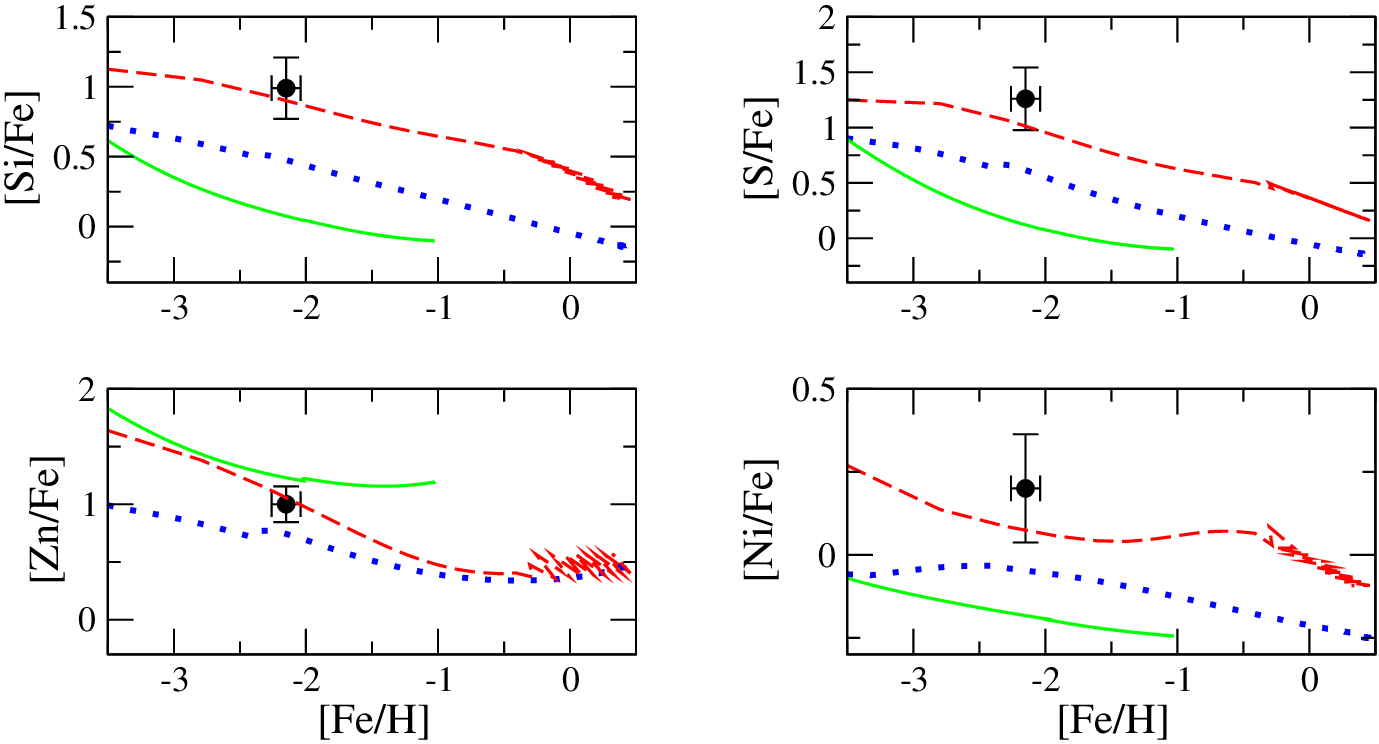}      
\caption{Observed abundance ratios vs. metallicity for the host galaxy of GRB 120815 provided by Kruhler et al. (2013).
The red-dashed (E01), blue-dotted (S00) and green-solid line (C00), are the predictions computed by means of the chemical evolution models for 
an elliptical, spiral and dwarf irregular galaxy, respectively. The star formation efficiency assumed in these models are
$\nu_{E01}=20 Gyr^{-1}$, $\nu_{S00}=1 Gyr^{-1}$, and $\nu_{C00}=0.1 Gyr^{-1}$.}
\label{120815}
\end{center}
\end{figure*}

Finally, in Figure \ref{120815}, we report the results for the host of GRB 120815,at $z=2.36$ observed by Kruhler et al. (2013); the elements analyzed here
are Si, S, Ni and Zn and all seem to indicate again a massive spheroid with strong SF.
In fact, to best fit the abundances in this host we adopted model E01 with SFE=20$Gyr^{-1}$, which is the maximum value we explored.  
It is interesting to note the high value of [Zn/Fe] which means that a higher quantity of dust is present along the line of sight of this GRB.

 In our analysis, although many uncertainties are still present in the adopted stellar yields and in the observational data 
and no firm conclusions on the GRB hosts can be drawn,
two of the three GRB host galaxies taken into account 
seem to favor either a spiral or a spheroidal host, while the third GRB (GRB120815) seems to be more confidently hosted by a proto-spheroid. 
This result, if confirmed in the future by more data, would be in contrast with
the claim that most of the HGs are small, star-forming and metal-poor galaxies
(see, e.g., Christensen et al. 2004, Fruchter et al. 2006 and Levesque et al. 2010).
Probably, the apparent discrepancy is due to the galaxy sample used to infer the
HGs properties, composed mainly of low redshift objects ($z \lesssim 2$). In particular,
recent works based on high redshift surveys seem to suggest a more complex view
(e.g. Kruhler et al. 2011; Hunt et al. 2011; Savaglio et al. 2012; Elliott et al. 2013,
Perley et al. 2013), where the majority of the HGs seem to suffer strong star
formation. This result is not unexpected recalling that GRBs are associated with
the death of massive and short-lived stars that trace the star formation inside the
host. The explanation lies in the evolution/changing of the star forming tracers
as a function of redshift: in the local universe the star formation is mainly traced
by small, blue galaxies, while at early epochs it is dominated by massive spheroids
and elliptical galaxies. Further evidence in favor of this hypothesis comes from
the study of the so called dark GRBs.
These GRBs are found at high redshift and they have in general a high metallicity. 
This is a very interesting fact suggesting that these GRBs arise from active
star formation in high redshift massive spheroids. Since ellipticals show in general
old stars, that means that we are catching ellipticals in formation. 
The characteristics behavior of [$\alpha$/Fe] ratios in spheroids shows a long plateau where they are
overabundant relative to Fe and the Sun.
 
This is due to the time-delay model which explains the behaviour of abundance ratios on the basis of the production of different elements in different stars 
with different timescales. In particular, $\alpha$-elements are mainly produced in CCSNe restoring them on short timescales, whereas Fe 
is produced mainly in Type Ia SNe and restored on a large range of timescales going from 30 Myr to a Hubble time (Matteucci 2001; 2012). This implies that, 
if the SFR is very efficient, the CCSNe can produce a substantial amount of Fe by themselves and when Type Ia SNe start restoring the bulk of Fe the 
abundance of this element in the ISM is already high. This produces high [$\alpha$/Fe] ratios at high metallicity. 
The contrary occurs when the SFR is quite low; the enrichment of Fe from core-collapse SNe is very modest and when Type Ia 
SNe start to restore the bulk of Fe, the abundance of this element in the ISM is still quite low. This produces low  [$\alpha$/Fe] ratios at low metallicity. 
Therefore, the diagrams [X/Fe] vs. [Fe/H] can be in principle very useful tools to infer the nature and the ages of unidentified objects.

\subsection{Age determination} \label{age}

In the chemical evolution models, the abundance ratios between two elements formed
on different timescales can be used as \textquotedblleft cosmic clocks\textquotedblright $\,$
and provide us with information
on the roles of LIMS and SNe in the enrichment process (Matteucci 2001).
As already mentioned, the study of abundance ratios such as [$\alpha$/Fe] is quite useful,
since the $\alpha$-elements are produced on short timescales by Type II SNe, whereas
the Fe-peak elements and nitrogen are produced on long timescales by Type Ia
SNe and low and intermediate-mass stars, respectively; we use the abundance
ratios of elements formed on different timescales to set important constraints on the age and the nature GRB HGs.

We define the age of the galaxy $A_{gal}$ as the time passed between the galaxy
formation epoch and the GRB observation time; this gives us a quantitative
measure of the time required for the galaxy to evolve and reach, at the time of
observation of the GRB, the chemical abundances observed during the afterglow
episode.
Practically, it is possible to infer it from the simultaneous comparison between
the [$\alpha$/Fe] vs [Fe/H] and the same ratio as a function of the galactic time, or
redshift, in a $\Lambda$CDM cosmological framework.
Applying the method to the three GRBs analyzed in this paper, we find
an age of 50 Myr for the galaxy hosting the GRB 120327A, 0.32 Gyr for the
GRB081008 host and 15Myr for the GRB120815 host.
These young ages indicate that we find GRBs to be "young" galaxies, namely galaxies in which the star formation 
is very active, probably suffering their first major star formation episode, but they are not necessary typical of 
the main population of galaxies that could have started forming stars already much before.

\section[]{The cosmic effective dust production rate} \label{cosmic}

The cosmic rate is defined in an unitary comoving volume of the Universe. 
This definition is necessary to study the rate at high redshift 
where the morphology of the observed galaxies is not known. The cosmic rate refers, 
in fact, to a mixture of galaxies which can be different at every redshift.
In particular, in order to evaluate the amount of dust observed at high redshift,
we have to take into account the cosmic dust cycle: dust is produced in stars and it is 
then blown off in a slow wind or a massive star explosion. The dust is then recycled in the clouds 
of gas between stars and some of it is consumed when the next generation of stars begins to form.

We compute the cosmic dust rate in units of 
$M_{\odot}\mbox{yr}^{-1} \mbox{Mpc}^{-3}$ by adopting the same method applied to the computation of
cosmic star formation rate, as  described in the paper of Grieco et al. (2012a). In that paper the cosmic SFR (CSFR) 
has been computed by taking into account the SF histories of galaxies of different morphological type as those shown in Fig. \ref{SFR}.

In particular the CSFR was defined as:
\begin{equation} \label{CSFR}
CSFR= \sum_{k}{\psi_{k}(t)\cdot n_{k}^{*}} \,\,\,(M_{\odot} yr^{-1} Mpc^{-3}), 
\end{equation}
where the SFRs have been then convolved with the number density for each type of galaxy. The quantity  $n_{k}^{*}$ is the galaxy 
number density, expressed in units of $\mbox{Mpc}^{-3}$, and $k$ identifies a particular galaxy type. In the Grieco et al. (2012a) paper, 
the main assumption was that the galaxy number density is constant in time and equal to the present time one for each galaxy type. 

Here we define, in the same way, the cosmic dust rate (CDR):

\begin{equation} \label{CDR}
CDR= \sum_{k}{[ (\dot {G}_{k}(t))_{prod}+(\dot {G}_{k}(t))_{accr}-(\dot {G}_{k}(t))_{destr} ] \cdot n_{k}^{*}} , 
\end{equation}
where $(\dot {G}_{k}(t))_{prod}$
is the total production rate of all the elements considered in the dust model (O, Mg, Si, S, Ca, Ni, Fe), and
$(\dot {G}_{k}(t))_{accr}$ and $(\dot {G}_{k}(t))_{destr}$ are the total accretion and destruction rates, respectively. 

In order to study different scenarios for the evolution of galaxy number density, here we assume:
$$n_{k}^{*}=n_{k,0} \cdot (1+z)^{\beta_k}$$
as suggested by Vincoletto et al. (2012). 
In particular, we explore two competitive scenarios:
\begin{enumerate}
 \item the {\itshape \textquotedblleft pure luminosity evolution\textquotedblright}, PLE, $\,$ corrisponding to the case proposed in the Grieco et al. (2012a);
 the CDR is obtained by assuming that all galaxies started forming stars at the same time and that there is no evolution in the galaxy number density from 
 early epoch to the present time. As a consequence, $\beta_{k}=0$ and $n_{k}^{*}=n_{k,0}$ for each galaxy type.
 \item The{\itshape \textquotedblleft density evolution\textquotedblright}$\,$ scenario, DE, where the $\beta$ parameter evolves with redshift, 
 as shown in Table \ref{CDRparam}, where the value of $n_{k,0}$, the same in both cases, and the values of $\beta_k$ for each galaxy type, are reported. 
 This particular combination of values is selected in order to obtain a CSFR which reproduces the trend of 
 a hierarchical galaxy formation scenario (e.g. Menci et al. 2004). In this picture, the majority of ellipticals formed in a large redshift range and 
 by mergers of spirals, with only a small percentage of them forming by means of a strong starburst at high redshift. 
 Clearly the CSFRs for the two scenarios are extreme cases, which predict completely different galaxy formation patterns 
 (see Fig. 6 of Grieco et al. 2012a, where both cases are shown). Because of this, we checked that the integrals of the CSFRs 
 in the two cases give a similar stellar density at the present time.

\end{enumerate}

\begin{table}
\begin{center}
\begin{tabular}{|c|c|c|}
 \hline
 \multicolumn{3}{c}{Density evolution scenario (DE)} \\
  \hline
 $\beta_E=-2.5$ & $\beta_S=+0.9$ & $\beta_I=0$ \\ 
  \hline 
  \multicolumn{3}{c}{  } \\
 \hline 
  \multicolumn{3}{c}{DE and PLE (in unit of $10^{-3}Mpc^{-3}$) } \\
 \hline 
  $n_{E,0}=4.4$ & $n_{S,0}=8$ & $n_{I,0}=0.2$ \\  
 \hline       
\end{tabular}
\end{center}
\caption{Parameters adopted for describing the pure luminosity evolution (PLE) and the density evolution (DE) cases. 
The quantities $n_{k,0}$ are expressed in $Mpc^{-3}$ and taken from Marzke et al. (1994).}
\label{CDRparam}
\end{table}

\begin{figure} 
\begin{center}
\includegraphics[scale=0.3]{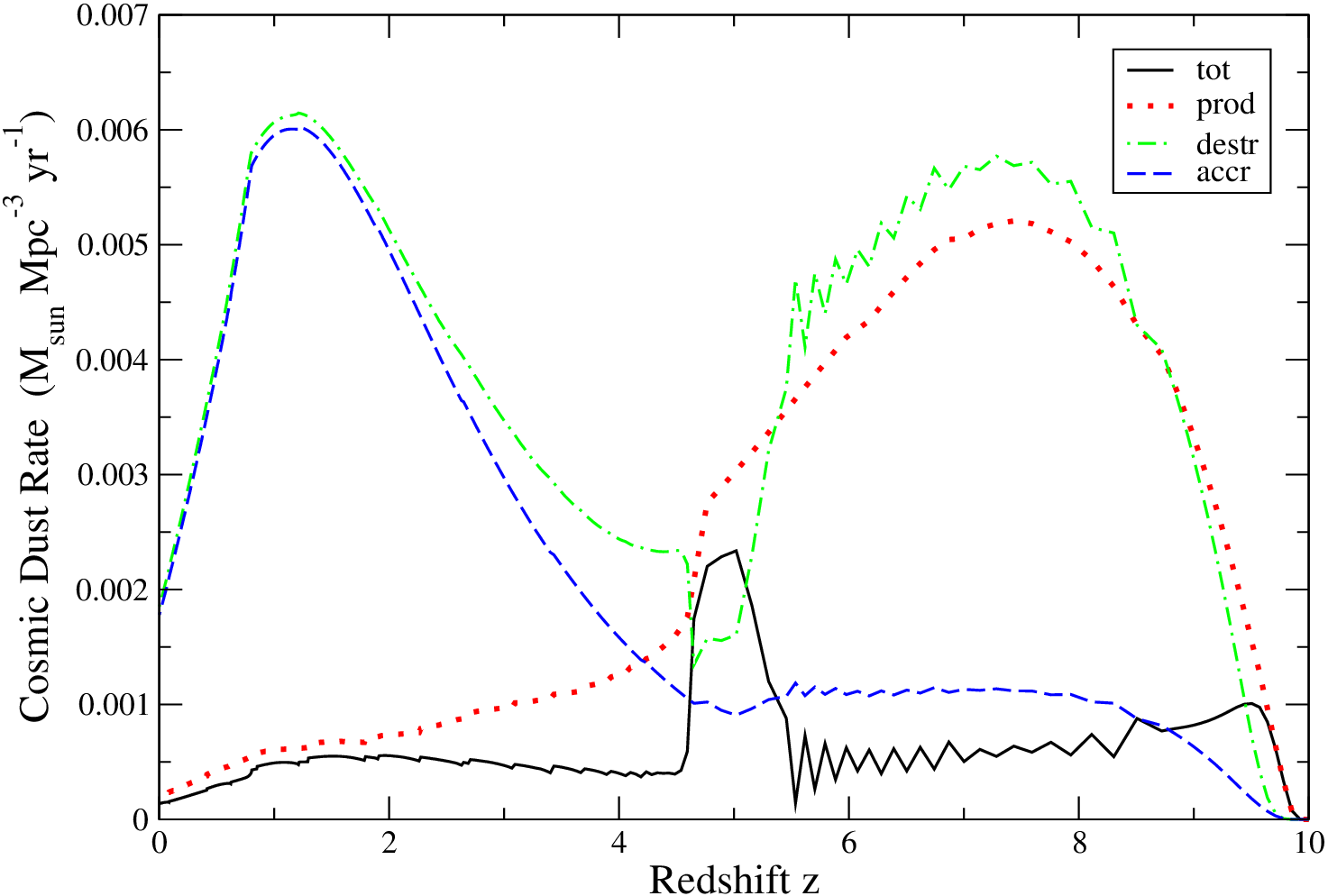}      
\caption{Cosmic effective dust production rate (black solid line), expressed as the sum of the cosmic stellar 
production (red dotted line), accretion (blue dashed line) and destruction (green dotted-dashed line) rates as a function of redshift, in the PLE scenario.}
\label{zCDustR}
\end{center}
\end{figure}

In Figure  \ref{zCDustR}, we show the predicted cosmic effective dust production rate as a function of redshift (black solid line), 
which includes production, accretion and destruction for each galaxy type.
This cosmic effective dust production rate, expressed in $M_{\odot} yr^{-1} Mpc^{-3}$, here has been computed by means of eq. \ref{CDR}, 
under the assumption of no number density evolution of galaxies. We remind that with no number density evolution we predict 
a very high star formation rate at high redshift due to the starbursts occurring during the early phases of the proto-spheroid formation 
(see Grieco et al. 2012a).
In the same plot are also shown, separately, the cosmic stellar dust production, dust accretion and dust destruction rates as functions of time. 
Fixing the redshift of galaxy formation at $z_f=10$, the predicted total cosmic effective dust production rate (black curve) shows a a 
peak at very high redshift followed by a 
decrease and then another higher peak at $z\sim5$.

\begin{figure} 
\begin{center}
\includegraphics[scale=0.3]{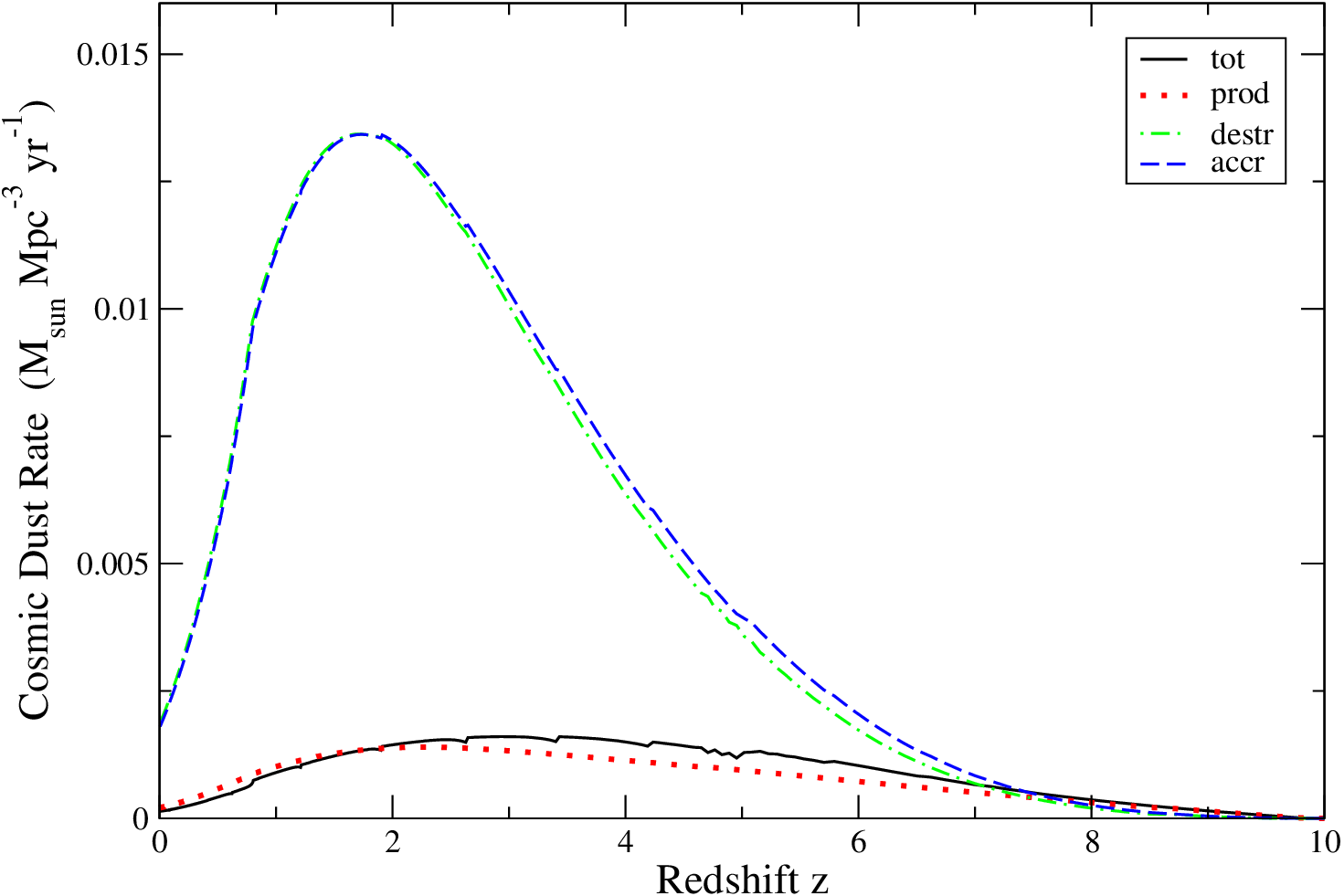}      
\caption{Cosmic effective dust production rate  as a function of redshift in the density evolution (DE) scenario. 
The lines have the same color set shown in Figure \ref{zCDustR}.}
\label{zCDR_DE}
\end{center}
\end{figure}

This behaviour is due to the predominance of proto-spheroids at high redshift where the early
high SN production induces first an increase in the dust rate but then a decrease due to the high destruction rate related to the SNe themselves. 
The peak at $z\sim 5$ is then due to the predominance of the star formation rate in the spirals, because since then, 
in this scenario, the ellipticals evolve passively.
The cosmic effective dust production rate then decreases strongly for $z <5$  down to redshift $z\sim 4.5$ and stays constant afterwards until $z=0$. 
It is interesting to note that accretion and destruction rates almost compensate each others during the whole cosmic time.
In Figure \ref{zCDR_DE}  we show instead the cosmic dust rates as functions of redshift in the case of strong number density evolution. 
It is interesting to note that the cosmic effective dust production rate increases slightly from high redshift 
down to $z\sim2$ and then decreases down to $z=0$. In general, the evolution of the  effective dust production rate is smoother than in the PLE case.
Clearly, the two different galaxy evolution scenarios produce different results concerning the cosmic dust evolution.

Finally, integrating in time the cosmic effective dust production rate, shown in Figures 8 and 9, we can obtain the cumulative cosmic dust 
mass density in the Universe as a function of redshift $\rho_{dust} (M_{\odot} Mpc^{-3})$.
In a $\Lambda$CDM cosmological frame with the critical density being:
$\rho_{crit}=\frac{3H_{0}^2}{8\pi G} \sim 1.36 \cdot 10^{11} \, M_{\odot} Mpc^{-3}$, for a Hubble constant of $70 km s^{-1} Mpc^{-1}$ we can 
compute $\Omega_{dust}$ and its evolution as a function of redshift, as shown in Figure 10 for the two scenarios.
The present time values of   $\Omega_{dust}$ from Figure 10 are:
$(\Omega_{dust})_{PLE}=\frac{\rho_{dust}}{\rho_{crit}} \sim 3.5\cdot 10^{-5}$ for the PLE scenario, and 
$(\Omega_{dust})_{DE}=\frac{\rho_{dust}}{\rho_{crit}} \sim 7.0\cdot 10^{-5}$ for the DE scenario.

Fukugita (2011) estimated a present time cosmic dust density produced by stars in the Universe of $\Omega_{dust} \sim 9.6 \cdot 10^{-6}$, 
whereas the value estimated from observations is $\sim 9 \cdot 10^{-6}$. 
Other estimates (Loeb \& Haiman 1997; Corasaniti 2007; Inoue \& Kamaya 2004; Fukugita 2011; Menard \& Fukugita 2012 ) 
all suggest a value between $10^{-5}$ and $10^{-6}$.
Our values are higher than these estimates and the reason could be that our stellar dust production is probably overestimated, 
in particular the assumed yield of dust from SNeIa could be overestimated, 
since there is no clear observational evidence for dust production in these SNe. However, we checked the effect of suppressing Type 
Ia SNe as dust producers and the effect on  $\Omega_{dust}$ is negligible.
On the other hand, it is reasonable that the observational estimate of the cosmic dust density is a lower limit. 
Finally, the fact that  $\Omega_{dust}$ is larger by a factor of two in the DE scenario is probably due to the fact that spiral galaxies 
dominate, at variance with the PLE scenario, where the number of spheroids is comparable to spirals. In the PLE scenario, the elliptical 
galaxies evolve very quickly and their star formation is quenched by galactic winds, whereas in the DE scenario ellipticals form by 
mergers of spirals where star formation continues until the present time and the dust production evolves more smoothly. 
Therefore, in the DE case more dust is expected. Moreover, a slightly higher cosmic density of stars is predicted for the DE scenario. 
We find:  $\Omega_{*}=0.005$ for the PLE case and  $\Omega_{*}=0.006$ for the DE case. In Fukugita \& Peebles (2004) 
this quantity is estimated to be  $\Omega_{*}=0.003$.

\begin{figure} 
\begin{center}
\includegraphics[scale=0.3]{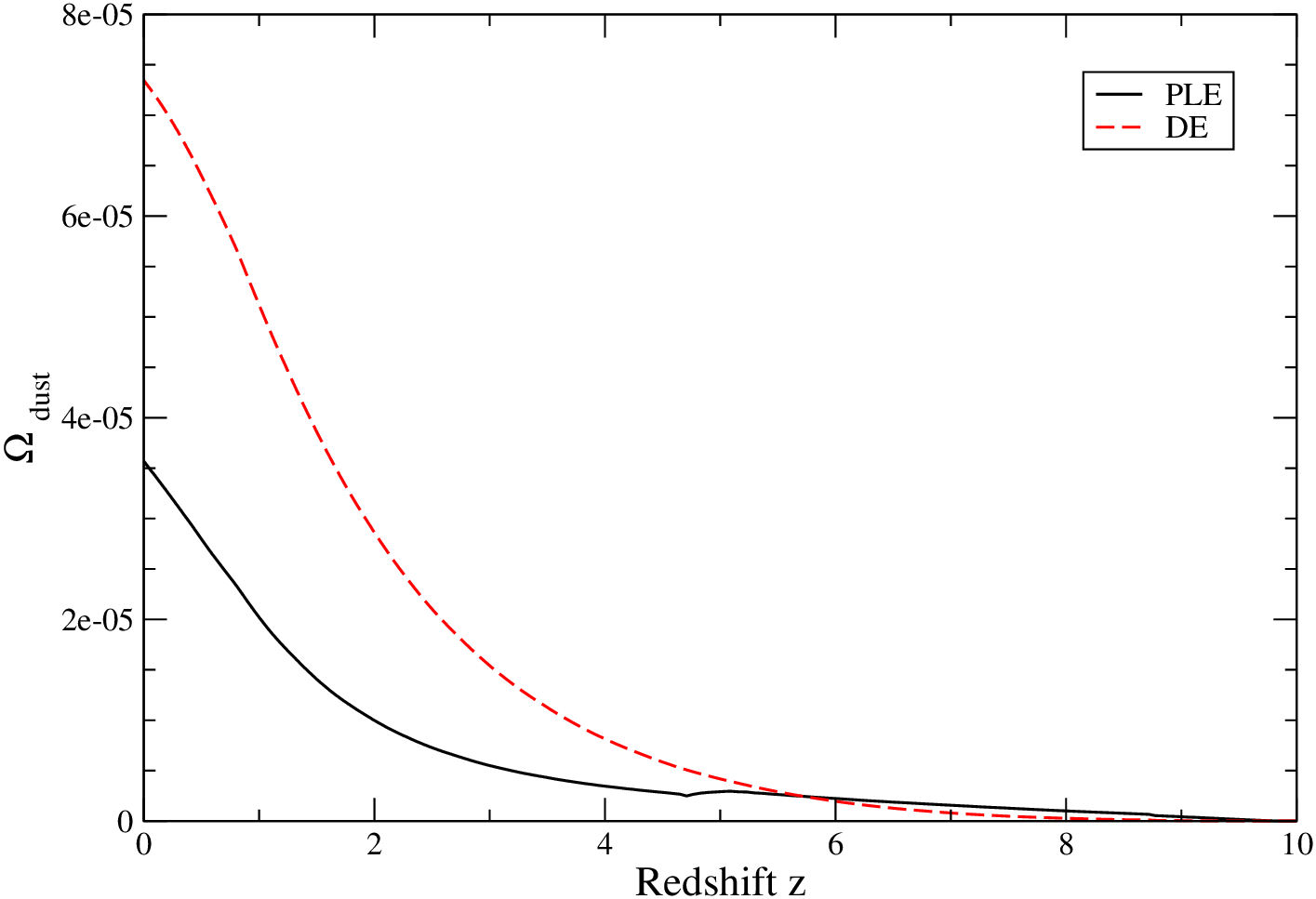}      
\caption{Predicted evolution of $\Omega_{dust}$ as a function of redshift in the PLE and DE scenarios. 
Continuous black line refers to the PLE scenario and the dashed red line the DE scenario.} 
\label{CumuCDRz}
\end{center}
\end{figure}

\section[]{Conclusions} \label{concl}
In this paper, we present a method, based on chemical evolution models, to identify the hosts of GRBs.  
Our idea  comes from the following considerations:
if GRBs are produced by the collapse of massive stars, they are presumably originating in galaxies where the bulk of massive star formation 
is taking place. From an observational point of view, almost in all the cases where the galaxy hosting the GRBs is observed, 
the burst position suggests that GRBs originate in regions where star formation is taking place (Bloom et al 2002).
In this context, the determination of the nature of GRB host galaxies and measurements of their star formation rate not only provide 
evidence in favour of or against the collapsar model, but also provide insight on the nature and origin of star formation in the early Universe.

We have presented a comparison between predicted and observed abundances and abundance ratios in some galaxies hosting GRBs 
(GRB081008, GRB120327A and GRB120815). 
To do that we have adopted detailed chemical evolutionary codes for galaxies of different morphological type including dust evolution.
The second aim of this study was to understand the effect of dust on the observed abundance patterns and to make predictions on the 
cosmic dust content as a function of the cosmic epoch.

Our main conclusions can be summarized as follows:
\begin{itemize}

\item We have studied the evolution of the abundances of C, N, O, Mg, Si, S, Ni and Zn in galaxies of different galaxies (ellipticals, spirals, irregulars).
Our results suggest that one of the galaxies hosting GRB120815 is probably a massive proto-spheroid catched during the active 
star formation phase, whereas for the other two galaxies hosting   
GRB120327A and GRB081008, no firm conclusion can be drawn, although some of the data seem to suggest spiral or proto-spheroid systems with high mass 
and star formation rate. In order to derive better conclusions we would need more precise stellar and dust yields. In particular, for some chemical 
elements the stellar yields are still quite uncertain (see Romano et al. 2010).
The result of  GRB120815 is particularly important since it is the first time that abundances measured in hosts of GRBs indicate early type galaxies.
Our result, if confirmed by more precise data,  suggest that GRBs can be  hosted in any type of galaxy 
suffering active star formation. Obviously, local spheroids which do not show any active star formation cannot host GRBs. We have also estimated the 
ages of the GRB hosts from chemical evolution and found that they are all young with ages ranging from 15 to 320 Myr.

\item We have shown that it is very important to include dust in the chemical evolution of galaxies although many uncertainties are still present in the 
recipes for the formation and destruction of dust. Here we have considered core-collapse SNe, AGB stars and Type Ia SNe. We stress that the dust is produced very early in 
the evolution of galaxies since CCSNe have lifetimes from few million to several tenths of million years and the massive AGB stars (7-8 $M_{\odot}$) have 
lifetimes of the order of 25-40 Myr.  Valiante et al. (2009) also showed  the importance of taking into account AGB stars to explain the dust in high redshift QSOs.
Another possible source of dust is the condensation
of dust grains in quasar winds (Elvis et al. 2002), which
has also been supported by observations (Markwick-Kemper et al. 2007). We did not include this source here since Pipino et al. (2011) 
have shown that it contributes negligibly to the total dust production in a massive elliptical, which is the typical QSO host.

\item We have computed the cosmic effective dust production rate, namely the rate of net dust production in an unitary volume of the Universe, 
by adopting the results of  our models and 
assuming either no number density evolution or strong number density evolution of galaxies, as suggested by the hierarchical clustering scenario 
for galaxy formation. They represent two extreme cases. 
The cosmic star formation rate in these two scenarios (see Grieco et al. 2012a) is different but the present time cosmic stellar density, which is 
the integral of the CSFR, is roughly the same. To compute the cosmic effective dust production rate, we took 
into account the rates of production from stars, destruction and accretion of dust in galaxies of different morphological type. 
We assumed as a redshift of galaxy formation $z_f=10$.
Our results show that in the pure luminosity evolution framework there are two peaks, one very close to the galaxy formation redshift and 
another stronger one at $z\sim5$. Then the dust rate decreases down to z=0. 
On the other hand, in the number density evolution framework the dust rate increases slightly from the beginning to $z\sim2$ 
and then decreases down to z=0.
 
\item Finally, we have estimated the cosmic dust density in the Universe at the present time, $\Omega_{dust}= \rho_{dust}/\rho_{crit}$ 
in both scenarios and the results are:
$\sim 3.5\cdot 10^{-5}$ in the case of pure luminosity evolution and $\sim 7\cdot 10^{-5}$ in the case of number density evolution of galaxies. 
Previous estimates (Loeb \& Haiman 1997; Corasaniti 2007; Inoue \& Kamaya 2004; Fukugita, 2011; Menard \& Fukugita, 2012) had 
suggested $\Omega_{dust}=10^{-6}-10^{-5}$. Our values are larger than previous estimates and the reasons for this can be found in
an overestimate of the dust production in our galaxy models due either to too large dust yields from stars (SNeII, Ia and AGB) or to the poor 
knowledge of the other processes invoving dust such as destruction and accretion. On the other hand, we cannot exclude that the amount of 
observed dust is underestimated. Therefore, these particular results should be taken with caution.

\end{itemize}

To conclude, it is worth noticing that in our analysis, two of the three GRB host galaxies taken into account are well fitted by a massive proto-spheroid model.
This result seems to be in contrast with the claim that most of the HGs are small, star-forming and metal-poor galaxies.
Probably, the apparent discrepancy is due to the galaxies sample used to infer the HGs properties composed mainly of low redshift objects (z $\lesssim$ 2).
% The key point to understand the whole picture is the star formation rate: the majority of the HGs in high redshift survey seem to have strong star formation while,
% as we already mentioned, at low redshift the observation point toward a population of star forming and metal poor galaxies and consequently with a milder star formation.
The explanation for this difference lies in the evolution/change of the star forming galaxies as a function of redshift:
in the local Universe the star formation is mainly traced by small, blue galaxies while at early epochs it is 
dominated by massive proto-spheroids.

 \section*{Acknowledgments}
V.G., F.M. and F.C., acknowledge financial support from PRIN MIUR2010-2011, project \textquotedblleft The chemical and dynamical 
evolution of the Milky Way and Local Group galaxies\textquotedblright, prot. N. 2010LY5N2T. V.G. and F.M. also acknowledge financial 
support from Trieste University through the Project FRA2011 on \textquotedblleft Cosmic star formation rate and cosmic chemical enrichment\textquotedblright. 
V.G. and F.M. thank A. Pipino and E. Spitoni for many useful discussions. 
Finally, we would like to thank an anonymous referee for his/her careful reading of the paper and very important suggestions.

\end{document}